\documentclass[aps,prd,nofootinbib,twocolumn,superscriptaddress,letterpaper,preprintnumbers, longbibliography]{revtex4-1}

\usepackage{amssymb}
\usepackage{amsmath}
\usepackage{epsfig}
\usepackage[colorlinks=true
,urlcolor=blue
,anchorcolor=blue
,citecolor=blue
,filecolor=blue
,linkcolor=red
,menucolor=blue
,hyperfootnotes=false,
,linktocpage=true
,pdfproducer=medialab
,pdfa=true
]{hyperref}
\usepackage{array}
\usepackage{booktabs}
\usepackage{comment}
\usepackage{cleveref}
\usepackage{multirow}

\newcolumntype{L}[1]{>{\raggedright\let\newline\\\arraybackslash\hspace{0pt}}m{#1}}
\newcolumntype{C}[1]{>{\centering\let\newline\\\arraybackslash\hspace{0pt}}m{#1}}
\newcolumntype{R}[1]{>{\raggedleft\let\newline\\\arraybackslash\hspace{0pt}}m{#1}}

\usepackage{afterpage}
\usepackage{longtable}
\usepackage{capt-of}

\usepackage{siunitx}
\DeclareSIUnit\electronvolt{e\kern-.05em V}
\DeclareSIUnit\parsec{pc}

\usepackage{graphicx}
\usepackage{gensymb}
\usepackage{subfigure}
\usepackage{blindtext}

\usepackage{listings}
\usepackage[dvipsnames]{xcolor}
\usepackage[T1]{fontenc}
\usepackage{fontawesome}

\definecolor{codegreen}{rgb}{0,0.6,0}
\definecolor{codegray}{rgb}{0.5,0.5,0.5}
\definecolor{codepurple}{rgb}{0.58,0,0.82}
\definecolor{backcolour}{rgb}{0.95,0.95,0.92}

\lstdefinestyle{mystyle}{
    language = Python,  
    commentstyle=\color{codegreen},
    keywordstyle=\color{magenta},
    keywordstyle=[1]\color[rgb]{0,0,0.75},
    keywordstyle=[2]\color[rgb]{0.5,0.0,0.0},
    keywordstyle=[3]\color[rgb]{0.127,0.427,0.514},
    keywordstyle=[4]\color[rgb]{0.4,0.4,0.4},
    commentstyle=\color[rgb]{0.133,0.545,0.133},
    numberstyle=\tiny\color{codegray},
    stringstyle=\color{codepurple},
    basicstyle=\ttfamily,
    breakatwhitespace=false,         
    breaklines=true,                 
    captionpos=b,                    
    keepspaces=false,                 
    numbers=none,                    
    numbersep=5pt,                  
    showspaces=false,                
    showstringspaces=false,
    showtabs=false,                  
    tabsize=2,
    morekeywords={True, False, len},
    columns=flexible
}

\newcommand\beq{\begin{alignat}{1}}
\newcommand\eeq{\end{alignat}}
\newcommand{\dhis}{\texttt{DarkHistory} }
\newcommand{\n}{\nonumber \\}

\newcommand*\bbar[1]{%
  \vbox{%
    \hrule height 0.5pt
    \kern-0.4ex
    \hbox{%
      \kern-0.2em
      \ifmmode#1\else\ensuremath{#1}\fi
      \kern-0.1em
    }
  }
}
\newcommand{\githubmaster}{\href{https://github.com/hongwanliu/DarkHistory/tree/lowengelec_upgrade}{\faGithub}\xspace}

\defcitealias{paperII}{Paper~II}

\begin{document}

\preprint{MIT-CTP/5523} 

\title{
	Exotic energy injection in the early universe I: \\a novel treatment for low-energy electrons and photons
}

\author{Hongwan Liu}
\email{hongwanl@princeton.edu}
\affiliation{Center for Cosmology and Particle Physics, Department of Physics, New York University, New York, NY 10003, U.S.A.}
\affiliation{Department of Physics, Princeton University, Princeton, New Jersey, 08544, U.S.A.}

\author{Wenzer Qin}
\email{wenzerq@mit.edu}
\affiliation{Center for Theoretical Physics, Massachusetts Institute of Technology, Cambridge, MA 02139, U.S.A.}

\author{Gregory W. Ridgway}
\email{gridgway@mit.edu}
\affiliation{Center for Theoretical Physics, Massachusetts Institute of Technology, Cambridge, MA 02139, U.S.A.}

\author{Tracy R. Slatyer}
\email{tslatyer@mit.edu}
\affiliation{Center for Theoretical Physics, Massachusetts Institute of Technology, Cambridge, MA 02139, U.S.A.}

\begin{abstract} 
	Decaying or annihilating dark matter and other exotic energy injections can modify the spectrum of the universe's photon bath, resulting in e.g. new contributions to spectral distortions of the cosmic microwave background blackbody spectrum and modifications to the temperature and ionization history of the universe.
	Here, we present an improved version of the \dhis code, which is now capable of consistently calculating the spectrum of low-energy photons by properly treating the interactions of these photons with the levels of hydrogen atoms.
	Other changes to the code include a more detailed treatment of energy deposition by low-energy electrons, and spectral distortions from heating of the intergalactic medium.
	All of the improvements we have made to \dhis are publicly available. \githubmaster
\end{abstract}

\maketitle

\section{Introduction}

The early Universe is an excellent laboratory for new physics searches. It was relatively homogeneous, making it simple to treat; between $20 \lesssim z \lesssim 150$, the intergalactic medium (IGM) temperature in standard $\Lambda$CDM cosmology evolves only through adiabatic cooling, making it exceptionally sensitive to exotic sources of heating; finally, the effect of new physics processes on the Universe can accumulate over a timescale at least a billion times longer than any feasible terrestrial probe of new physics. As a result, energy injection from new physics that is otherwise undetectable terrestrially or in our local astrophysical neighborhood can both be accurately predicted and potentially observed with early-Universe probes.

High-energy Standard Model (SM) particles injected from new physics processes lose their energy as they interact with the IGM, ultimately depositing their energy into the following channels: 
\textit{1)} ionization, resulting in an increase in the free electron fraction $x_e \equiv n_e / n_\text{H}$, where $n_e$ and $n_\text{H}$ are the number densities of free electrons and hydrogen (both neutral and ionized) respectively; 
\textit{2)} heating of the IGM, resulting in an increase in the IGM temperature $T_m$; 
\textit{3)} the production of low-energy photons, a term we use to refer to photons below the ionization potential of hydrogen, $\mathcal{R} \equiv \SI{13.6}{\eV}$;
and potentially \textit{4)} the increased abundance of excited states of atoms.
Low-energy photons can show up as extragalactic background photons over a wide range of frequencies.\footnote{For generic unstable Standard Model particles, a significant portion of their energy can go into free-streaming neutrinos as well, but we will only focus on electromagnetic channels in this work.} 

Searches for new physics in all three of these channels have been studied extensively in the literature. The increase in ionization levels from dark matter (DM) annihilation and decay, and its impact on the cosmic microwave background (CMB) anisotropy power spectrum, provides some of the strongest limits on the rates of such processes for sub-GeV dark matter~\cite{0906.1197, 0907.3985, 1506.03811, 1610.06933, 1610.10051,Cang:2020exa}. 
Exotic heating of the IGM during the cosmic dark ages could modify the 21-cm brightness temperature significantly, so that future 21-cm observations could potentially lead to strong constraints on DM annihilation, decay, and primordial black holes (PBHs)~\cite{1408.1109, 1603.06795, 1803.03629, 1803.09739, 1803.09398, 1803.11169, 1803.09390}. Heating of the IGM during reionization can also be probed by temperature measurements of the IGM using the Lyman-$\alpha$ forest~\cite{1710.00700, 1808.04367, 2001.10018, 2009.00016}, setting constraints on DM velocity-dependent annihilation and decay~\cite{0907.0719, 1308.2578, 1604.02457, 2008.01084}; dark photon dark matter can likewise convert into extremely low-frequency photons that heat the IGM~\cite{1911.05086, 2002.05165, 2003.13698}, and may potentially provide sufficient heating to reconcile low- and high-redshift Lyman-$\alpha$ observations of the IGM~\cite{2206.13520}. 
Finally, the limits on CMB spectral distortions have been used to constrain dark matter scattering against Standard Model particles~\cite{Ali-Haimoud:2015pwa}.
In addition, there have been a number of studies that employ the method of Green's functions to study spectral distortions from heating~\cite{Chluba:2013vsa}, photon injection~\cite{Chluba:2015hma}, and electron injection~\cite{Acharya:2018iwh} prior to recombination.

The public \dhis code~\cite{DH} was designed to compute changes to the cosmic temperature and ionization history due to exotic energy injections, self-consistently taking into account energy injection from conventional sources (e.g. stars during the epoch of reionization).
In this paper, we make significant improvements to the computation of the global free-electron fraction $x_e$, IGM temperature $T_m$, and the intensity of low-energy photons $I_\omega$ 
\footnote{Throughout this work, we will use angular frequency $\omega$ instead of frequency $\nu$, in accordance with our choice of natural units.}
in the presence of exotic energy injection, focusing on injection occurring during or after recombination, and provide publicly available tools for these computations.~\footnote{Our code is available at \githubmaster.}
The key advances are: 
\begin{enumerate}
    \item We improve our treatment of electron cooling by making significant upgrades to our treatment of low-energy ($< \SI{3}{\kilo\eV}$) electrons---especially in terms of collisional excitation into a range of hydrogen excited states---and inverse Compton scattering. This in turn allows us to track low-energy ($< \SI{10.2}{\eV}$) photons produced during electron cooling;
    \item We account for $y$-type spectral distortions produced by heating of baryons; 
    \item We track the population of a range of excited states in the hydrogen atom, as well as transitions between these states. We also allow low-energy photons produced from exotic energy injection to excite and de-excite hydrogen atoms, while also carefully tracking the low-energy photon spectrum produced by excitation and de-excitation. 
\end{enumerate}

Put together, these improvements result in two major achievements. 
First, they allow us to consistently calculate, for the first time, the frequency-dependent change in the photon background intensity $\Delta I_\omega$ over the CMB after recombination due to exotic sources of energy injection, enabling the use of spectral distortions to the CMB to look for such processes. 
Second, the occupation number of photons injected by dark matter with energies between the CMB temperature after recombination and the ionization potential of hydrogen significantly exceeds the occupation number of the CMB, and may lead to corrections to the ionization and thermal histories.
The ability to track $\Delta I_\omega$ enables us to quantify the influence of sub-\SI{10.2}{\eV} low-energy photons on the process of recombination, improving the reach of energy injection calculations to lower energies, as well as increasing the accuracy.

In a companion work, which we call \citetalias{paperII}, we explore the applications of our improved calculation~\cite{paperII}. 
We provide the full space of predicted low-energy photon spectra---from radio to ultraviolet frequencies---for dark matter annihilation and decay into electron/positron pairs and photons, providing a new-physics benchmark for future CMB spectral distortion experiments. 
We show the corrections imparted by additional low-energy photons (injected by dark matter annihilation and decay) on the ionization and thermal histories. These corrections are at the level of a few percent; hence any results using the ionization histories calculated with \dhis \texttt{v1.0} are largely unchanged. 
We also show for the first time the ionization histories resulting from dark matter decaying to photons with masses less than \SI{10}{\kilo\eV}.
Finally, our improved treatment of low-energy electrons also allows us to extend existing CMB anisotropy power spectrum limits on the annihilation and decay of DM into photons to DM masses of less than \SI{10}{\kilo\eV}, giving rise to important limits on the coupling between axion-like particles (ALPs) and photons in this mass range that we present in \citetalias{paperII}. 

The complete contribution to the low-energy photon spectrum during and after recombination including exotic energy injection will be computed here for the first time.
In Section~\ref{sec:lowengelec}, we describe our improved treatment for low-energy electrons, including the ability to now track the spectrum of photons resulting from inverse Compton scattering (ICS) by these electrons. 
In Section~\ref{sec:y-type}, we describe the contributions of ICS and heating of the IGM to the low-energy photon spectrum.
Section~\ref{sec:evolution} outlines our method for evolving low-energy photons, by taking the contributions mentioned above and including their interactions with hydrogen atoms.
We conclude with Section~\ref{sec:conclusion}.
In addition, we include a number of appendices which provide more detailed discussion to support the main text.
Throughout this work, we will use natural units, where $c = \hbar = k_B = 1$.

\section{Low-Energy Electrons}
\label{sec:lowengelec}

In this section we describe how an injected electron or positron loses its energy in the early universe as it cools. 
Since much of the following discussion builds directly on Ref.~\cite{DH}, 
we focus here on extensions and improvements we have made. 
In Ref.~\cite{DH}, electrons were artificially divided into high-energy ($> \SI{3}{\kilo\eV}$) and low-energy electrons ($\leq \SI{3}{\kilo\eV}$). 
For high-energy electrons, the particles undergo all possible cooling processes with some probability, and one can therefore solve a set of linear equations to determine the energy deposited by each process.
Previously in \dhis \texttt{v1.0}, when the electrons cool to kinetic energies below $\SI{3}{\keV}$, the interpolation tables provided by Ref.~\cite{MEDEAII} determined where the remaining energy was ultimately deposited. 
With this method, it is possible to track the total energy converted into photons with energy less than $E_\alpha =$ \SI{10.2}{\eV}, but not the \textit{spectrum} of these photons. 
In the subsequent section, we present the following changes to \dhis:
\begin{itemize}
	\item We update our collisional excitation cross-sections and extend our treatment of high-energy electrons to all electrons.
	\item In addition to tracking how low-energy electrons deposit their energy, we now simultaneously track the spectrum of photons they produce as they cool.
\end{itemize}

\subsection{Energy deposition from low-energy electrons}
\label{sec:eng_dep}

Electrons of all relevant energies injected into the Universe after recombination cool extremely efficiently, losing almost all of their energy by scattering off atoms, ions, electrons, and CMB photons over a timescale that is much shorter than a Hubble time~\cite{0906.1197}. 
Consider an electron produced with energy $E'$. 
We would like to know how much energy this electron deposits, $R_c(E')$, into the IGM through a given channel $c$. 

As explained in Ref.~\cite{DH}, $R_c(E')$ can be calculated by considering the electron to cool over a timescale much shorter than the characteristic timescales of all cooling processes, and calculating the energy deposited promptly into all channels $c$, as well as the secondary electron spectrum $d N_e / dE$. Then $R_c(E')$ can be calculated recursively using the integral equation
\begin{alignat}{1}
    R_c(E') = \int dE \, R_c(E) \frac{d N_e}{dE} + P_c(E') \,,
    \label{eqn:elec_cooling_analytic}
\end{alignat}
where $P_c(E')$ is the energy promptly deposited into channel $c$. A similar equation can be written for the spectrum of photons produced through ICS. 
When discretized, this equation becomes a lower-triangular system of linear equations that we can solve numerically.

Previously in \dhis \texttt{v1.0}, this formula was only applied to ``high-energy electrons'' with $E'\geq\SI{3}{\keV}$, resolving them into (1) what was termed ``high-energy deposition'' with channels ionization, Lyman-$\alpha$ excitation, and heating, (2) a spectrum of ICS photons that was passed to the photon cooling part of the code, and (3) a low-energy $E'<\SI{3}{\keV}$ electron spectrum. 
This treatment was not extended to ``low-energy electrons'' due to approximations made in the cross-sections used in \dhis \texttt{v1.0}  that may not apply at lower energies. Instead, the energy deposited by these low-energy electrons was resolved by default using the results of the MEDEA code~\cite{MEDEAII}, which studied the cooling of low-energy electrons using Monte Carlo methods, ultimately resolving electrons into hydrogen ionization, helium ionization, Lyman-$\alpha$ excitation, heating, and continuum (sub-$E_\alpha$) photons for a range of different ionization levels, $x_e$. 
This division was also intended to make it easier for users to incorporate their own alternative models for the cooling of low-energy electrons (e.g.~accounting for updated cross-sections, the inclusion of additional states, or more sophisticated methodology) into \texttt{DarkHistory}, without needing to modify the rest of the code.

However, this division by energy is not a requirement;
 Eq.~\eqref{eqn:elec_cooling_analytic} applies to all electrons, regardless of their energy, and we can simply extend the previous high-energy electron treatment to cover all regimes so long as we have cross sections that apply across all relevant energies. 
 We note that the authors of Refs.~\cite{Acharya:2019uba} and~\cite{Jensen:2021mik} also treated electrons with no such division between high and low energies. 
 In our current work, we have updated the collisional ionization and excitation cross-sections of electrons with HI, HeI and HeII. For collisional ionization, we adopt the binary-encounter models described in Ref.~\cite{Kim_Rudd}, while for collisional excitation, we use results from Ref.~\cite{Stone_Kim_Desclaux} for all hydrogen $np$ states and HeI, and the CCC database~\cite{CCC} for all other hydrogen states. These results are extended to higher energies using the Bethe approximation~\cite{RevModPhys.43.297}. Further details on the cross sections can be found in Appendix~\ref{app:collisional_rates}. 

With these new cross sections, we now calculate the energy deposition of both low- and high-energy electrons using Eq.~\eqref{eqn:elec_cooling_analytic}, and extend the number of possible excitation transitions from the hydrogen and helium ground states to cover the following states: 
\begin{itemize}
	\item H excitation from $1s$ to any $nl$ state with $n\leq 4$, where the integers $n,l$ index the principal quantum number and angular momentum quantum number of the hydrogen atom,
	
	\item H excitation from $1s$ to any $np$ state with $n\leq 10$,
	
	\item HeII excitation to the first excited state and HeI excitation to the $n^{2S+1}L = 2^3s$ state. Here, $S$ and $L$ are the total spin and orbital angular momenta, respectively.
\end{itemize}
The default treatment in \dhis now includes all of these states, but users can change this as desired. 

At the end of our electron cooling calculation, the initial electron energy is subdivided between heating of the IGM, ionization of neutral hydrogen, ionization of neutral and singly ionized helium, excitation of neutral hydrogen, excitation of neutral helium, and a spectrum of photons. 
The spectrum of photons produced by electron cooling is then added to the propagating photon spectrum at the current timestep, and is resolved in the same manner as in \dhis \texttt{v1.0}. 
However, the excited states should also ultimately cascade down to the ground state, producing more photons that need to be treated as well. 
We have developed two methods to treat these deexcitation photons, under the simplifying assumption that deexcitation from helium can be neglected, given that the number density of helium nuclei $n_\text{He}$ is only 8\% of $n_\text{H}$. 

The first method involves tracking the population of a large number of excited states using the multilevel atom method described in Ref.~\cite{preHyrec}, and provides an accurate way of tracking photons that are emitted after deexcitation; we will describe this method in detail in Section~\ref{sec:evolution}. 
For calculations involving e.g. CMB spectral distortions, it is necessary to use this method.

The second method follows the prescription used in MEDEA~\cite{MEDEAII}; for any calculations that do not require accurately tracking the background spectrum of photons, this method is sufficient and faster than the first method.
Specifically, every excitation to an $nl$ state cascades down to either the $2s$ or the $2p$ state, emitting sub-$E_\alpha$ photons in the process. 
If the electron reaches the $2s$ state, it undergoes a two-photon forbidden decay process, further emitting sub-$E_\alpha$ photons; otherwise, the excitation is counted as an effective Lyman-$\alpha$ excitation. 
The probability for an $nl$ state cascading down to a $2s$ or $2p$ state is calculated in Ref.~\cite{Hirata:2005mz}. 
In this way, the initial energy of an electron is completely distributed among the following channels: \textit{1)} ionization of neutral hydrogen; \textit{2)} ionization of neutral and singly-ionized helium; \textit{3)} Lyman-$\alpha$ excitation; \textit{4)} heating of the IGM, and \textit{5)} sub-$E_\alpha$ continuum photons. 
We have also implemented this second method in \dhis and will use this method to compare our results with the existing literature in Section~\ref{sec:cross-checks}. 

\subsection{ICS of low-energy electrons}
\label{sec:ICS}

In addition, we obtain the resultant low-energy photon spectrum, $N^\text{ICS}_\omega$, produced during the cooling process through inverse Compton scattering of CMB photons. 
As explained in Ref.~\cite{DH}, we can use a similar strategy as in Section~\ref{sec:eng_dep} to determine the CMB distortion caused by ICS. 
The distortion is defined as the upscattered photon spectrum, subtracting off the original CMB photons that were scattered.
Previously, this method was applied to high-energy electrons, but ICS was neglected for electrons with energy less than 3 keV, since ICS is a subdominant process for these energies after recombination.
Now we extend the treatment to arbitrarily low energies. 

For low enough electron energies, the amount of energy gained by the upscattered CMB photons scales as $\beta^2$, where $\beta$ is the electron velocity; hence the upscattered photon spectrum becomes very similar to the original CMB spectrum as the electron's energy becomes small.
This leads to catastrophic cancellation between the two terms when determining the CMB distortion.
To solve this issue, we use an analytic formula for the difference of the two terms.

At low energies, the spectrum of photons produced per unit time due to ICS of low-energy electrons can be given by a Taylor expansion in $\beta$,
\begin{multline}
\frac{dN^\text{ICS}_\omega}{dt} = n_\text{CMB}(\omega, T) \sigma_T \\
+ \frac{3 \sigma_T T^2}{32 \pi^2} \sum_{n=1}^\infty \sum_{j=1}^{2n}\frac{A_n \beta^{2n} x^3 P_{n,j}(x) e^{-jx}}{(1 - e^{-x})^{2n+1}} \, ,
    \label{eqn:thomson_scattered_phot_spec_expansion}
\end{multline}
where $N^\text{ICS}_\omega$ is the number of upscattered photons per energy $\omega$, $n_\text{CMB}(\omega, T)$ is the blackbody spectrum (number density of photons per unit energy), $\sigma_T$ is the Thomson cross-section, $A_n$ is a constant, $x \equiv \omega/T$, and $P_{n,j}(x)$ is a rational or polynomial function in $x$. 
Expressions for $A_n$ and $P_{n,j}(x)$ are given in the appendix of Ref.~\cite{DH}. 
The first term in the expansion is exactly the original spectrum of CMB photons that were upscattered. 
This is precisely the expected result as $\beta \to 0$, since we approach the Thomson scattering limit and photons are simply scattered elastically. 
Terms that are higher order in $\beta$ can therefore be regarded as distortions to the blackbody spectrum that occur due to scattering, and the energy loss can be directly computed from those terms alone. 
In fact, we can check that when integrated over $\omega$, 
\begin{multline}
  \int d\omega \sum_{n=1}^\infty \sum_{j=1}^{2n} \omega \frac{A_n \beta^{2n} x^3 P_{n,j}(x) e^{-jx}}{(1 - e^{-x})^{2n+1}} \\
  = \frac{4}{3} \sigma_T c \beta^2 (1 + \beta^2 + \beta^4 + \beta^6 + \cdots) u_\text{CMB}(T) \,,
\end{multline}
where $u_\text{CMB}(T)$ is the blackbody energy density with temperature $T$; this is simply the Taylor expansion of the energy loss rate of an electron scattering off a blackbody spectrum with temperature $T$ in the Thomson regime, $dE' / dt = (4/3) \sigma_T \beta^2 \gamma^2 u_\text{CMB}(T)$, where $\gamma \equiv (1 - \beta^2)^{-1/2}$. 
Therefore, we compute the distortion produced by low-energy electrons directly by computing only the $n \geq 1$ terms in Eq.~\eqref{eqn:thomson_scattered_phot_spec_expansion}; in practice, we include terms up to $n = 6$.

\subsection{Comparison to other calculations}
\label{sec:cross-checks}

%
\begin{figure*}
	\includegraphics[scale=0.45]{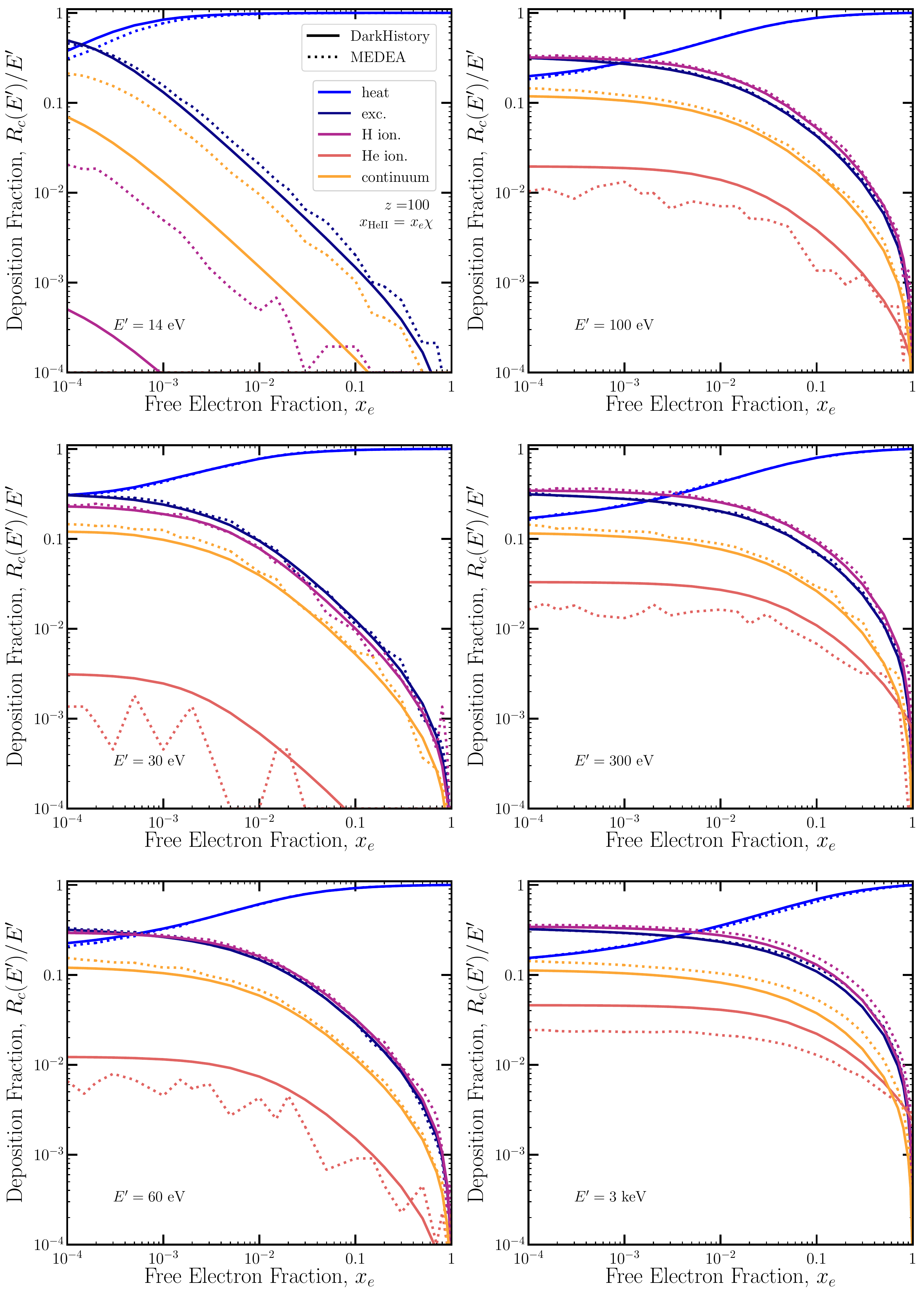}
	\caption{
	A comparison of the fraction of energy deposited into various channels by low-energy electrons, $R_c (E') / E'$, as calculated by \dhis (solid) and Ref.~\cite{MEDEAII} (dashed), where $E'$ is the energy of the electron. 
	The channels we include are heat (blue), Lyman-$\alpha$ excitation (indigo), hydrogen ionization (magenta), helium ionization (orange), and sub-$E_\alpha$ or `continuum' photons (yellow). 
	}
	\label{fig:MEDEA_xcheck}
\end{figure*}
\begin{figure*}
    \includegraphics[scale=0.45]{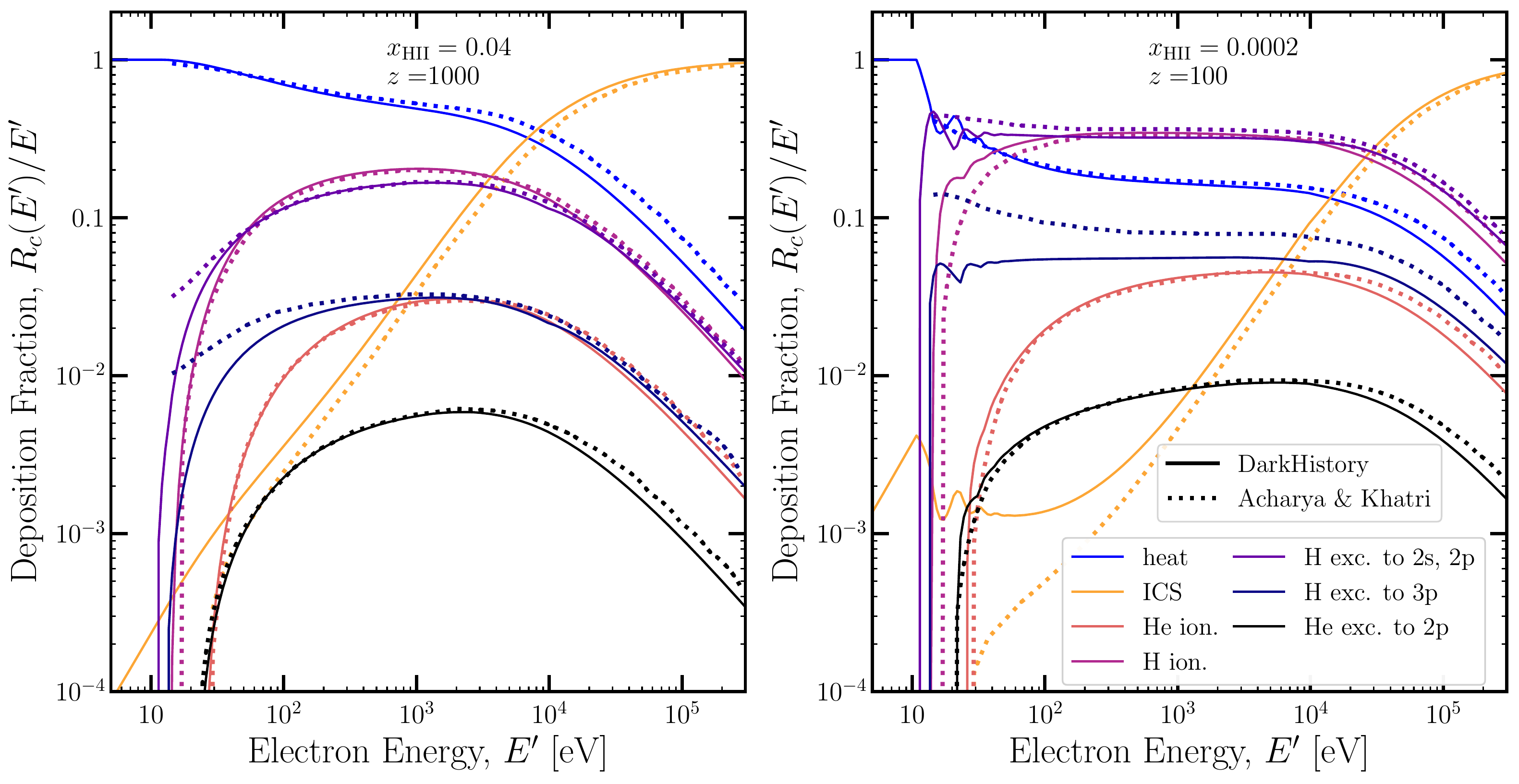}
	\caption{
	A comparison of the fraction of energy deposited into various channels by low-energy electrons, $R_c (E') / E'$, as calculated by \dhis (solid) and Ref.~\cite{Acharya:2019uba} (dashed). 
	The channels we include are heat (blue), ICS (yellow), helium ionization (orange), hydrogen ionization (magenta), 
	hydrogen excitation to the $n=2$ (purple) and $n=3$ state (indigo), and helium excitation to the $n^{2S+1}L = 2^1p$ state (black). 
	}
	\label{fig:AcharyaKhatri_xcheck}
\end{figure*}

In this section we compare our calculation of $R_c(E')$ to that of the MEDEA code~\cite{MEDEAII} and Ref.~\cite{Acharya:2019uba}. 
The former implements a Monte Carlo calculation of $R_c(E')$ while the latter solves a lower-triangular system of equations much like we do. 

To make the comparison with MEDEA, we use the second method described at the end of Section~\ref{sec:eng_dep}, restricting ourselves only to the $2s$ and $np$ states, in agreement with their method. 
In Fig.~\ref{fig:MEDEA_xcheck} we compare our results against the output of the MEDEA code~\cite{MEDEAII}, summarized in Ref.~\cite{Galli:2013dna}. 
Following Ref.~\cite{MEDEAII}, we plot $R_c(E') / E'$ against the free electron fraction $x_e$ for six different values of the input electron energy at $z=0$, and set the helium ionization level to linearly scale with the hydrogen ionization level, $x_\text{HeII} = x_\text{HII} \, \chi$, where $\chi \equiv n_\text{He}/n_\text{H}$ is the ratio of all helium to hydrogen nuclei by number. 

We find generally good agreement between the two methods.
However, compared to our method, relying on the MEDEA results has two disadvantages. First, Monte Carlo methods generally struggle to capture rare channels and continuous processes properly. 
The inability to capture rare channels accurately is evident in the noticeable differences between the helium ionization curves; our results are smoother across $x_e$ and tend to be slightly larger in amplitude.
Our \SI{14}{\eV} result also differs significantly from MEDEA's results, predicting a larger contribution to heating, and a suppression to energy deposited into continuum photons and hydrogen ionization (helium ionization is impossible for such low-energy electrons). This may be due to the fact that MEDEA models the heating loss process as discrete events, each leading to a loss of 5\% of the total kinetic energy of the electron. Since \SI{14}{\eV} electrons are close to the hydrogen ionization threshold, this discretization of a continuous process may be the cause of the discrepancy.  
The second disadvantage in relying on the MEDEA results is that we only have data for electrons with 14, 30, 60, 100, 300, and 3000 eV of kinetic energy; an interpolation must be performed over the kinetic energy for all other values. Our new method avoids this problem, and is able to compute $R_c(E')$ at any arbitrary energy. 

In Fig.~\ref{fig:AcharyaKhatri_xcheck} we compare our results for $R_c(E')$ against those of Ref.~\cite{Acharya:2019uba}. 
Again, the overall agreement is generally good, though there are a few qualitative differences. 
We find that the two results sometimes differ at excitation thresholds, with our results going smoothly to zero as energy decreases. This may be due to a difference in resolution for $E'$.

Additionally, some of our curves have oscillatory features while those of Ref.~\cite{Acharya:2019uba}  do not.
These oscillatory features are present in e.g. Refs.~\cite{1979ApJ...234..761S,2010MNRAS.404.1869F} and have a simple explanation. 
Consider the hydrogen excitation to the $n=2$ state in the right hand panel. 
Each minimum corresponds to a multiple of the Lyman-$\alpha$ energy $E_\alpha$. 
If the injected electron has energy just under $2 E_\alpha$, then after it excites an atom, it will have just under $E_\alpha$ energy left over, and therefore is unable to excite any more atoms.
As the injected electron energy increases above this threshold, then the secondary electrons will be energetic enough to also deposit energy into excitation.
The suppression of energy going into excitation means that energy must be deposited through other processes, hence we see maxima at kinetic energies of $2 E_\alpha$ in other channels.
The oscillations at the other multiples of $E_\alpha$ have a similar explanation. 

Finally, for completeness, we note that Ref.~\cite{Jensen:2021mik} proposed an analytic method to estimate the fraction of electron energy deposited directly into continuum photons; for sufficiently low-energy electrons, the remaining energy goes into heating, ionizations, and excitations. 
Within its regime of validity, this approach agrees very well with the results of \dhis \texttt{v1.0} presented in Ref.~\cite{DH}.

\section{Spectral distortions from Compton scattering}
\label{sec:y-type}

In addition to the ICS secondary photons described in Section~\ref{sec:ICS}, heating of the IGM from exotic sources also leads to a distortion of the CMB away from the blackbody spectrum, as photons Compton scatter off a thermal distribution of electrons at a different temperature. 
For redshifts $z \lesssim 5 \times 10^4$, photon energies can no longer be efficiently redistributed to establish a Bose-Einstein distribution, leading to a $y$-distortion~\cite{Zeldovich_Sunyaev}. 
The shape of the $y$-type distortion is given by 
\begin{equation}
\Delta I_\omega = y \times \frac{\omega^3}{2 \pi^2} \frac{x e^x}{(e^x - 1)^2} \left[ x \coth \left( \frac{x}{2} \right) - 4 \right] .
\label{eqn:y-distortion}
\end{equation}
The amplitude of this distortion is controlled by the $y$-parameter, which is given by~\cite{Chluba:2018cww}
\begin{equation}
y = \int_0^t  dt \, \frac{T_m - T_\text{CMB}}{m_e} \sigma_T n_e \,,
\label{eqn:y-param}
\end{equation}
where $T_\text{CMB}$ is the CMB temperature and $m_e$ is the electron mass.
As defined, $y$ includes contributions from \textit{1)} the adiabatic cooling of baryons, which causes $T_m < T_\text{CMB}$ at $z \lesssim 155$; \textit{2)} photoheating during the process of reionization, and \textit{3)} exotic sources of energy injection. 
We define the contribution to $y$-distortions just from exotic energy injection as $y_\text{inj}$, which can be parametrized as
\begin{equation}
    y_\text{inj} = \int_0^t dt\, \frac{\Delta T}{m_e} \sigma_T n_e  \,, \qquad \Delta T = T_m - T_m^{(0)},
    \label{eqn:y_DM}
\end{equation}
where $T_m^{(0)}$ is the temperature history including reionization in the absence of exotic energy injection.
Throughout this paper, we assume the default reionization model used in \dhis \texttt{v1.0}, which is based on Ref.~\cite{Puchwein:2018arm}; users may define their own reionization model in performing these calculations. 

At early enough times, Eq.~\eqref{eqn:y_DM} reduces to an integral over the heating rate, a form that is useful numerically when baryons are tightly coupled to the CMB with $T_m \approx T_\text{CMB}$, and is typically seen in the literature when discussing spectral distortions due to energy injection before recombination.
We give a novel derivation of this expression in Appendix~\ref{app:y_deriv}, following a similar calculation from Ref.~\cite{Hirata:2008ny}, and present the main results here.
At redshifts well before reionization, the evolution equation for the matter temperature is given by~\cite{DH}
\begin{equation}
	\dot{T}_m = -2 H T_m + \Gamma_C (T_\text{CMB} - T_m) + \dot{T}_m^\text{inj} ,
	\label{eq:general_Tm_evolution}
\end{equation}
where $H$ is the Hubble parameter, and $\Gamma_C$ is the Compton heating rate coefficient. The rate of change of the matter temperature due to exotic energy injection is defined by
\begin{alignat}{1}
    \dot{T}_m^\text{inj} \equiv \frac{2 \dot{Q}}{3 (1 + \chi + x_e) n_\text{H}} \,, \quad \dot{Q} \equiv f_\text{heat} \left(\frac{dE}{dV \, dt} \right)_\text{inj} \!\!\!.
\end{alignat}
$\dot{Q}$ is the usual exotic heating rate in terms of energy density per time and is parameterized in terms of the energy injection rate using the coefficient $f_\text{heat}$.
Define the quantity 
\begin{equation}
J \equiv \frac{8 \sigma_T u_\text{CMB} x_e}{3 (1 + \chi + x_e) m_e H} \,.
\end{equation}
The integrated $y$-parameter as defined in Eq.~\eqref{eqn:y-param}, including the $\Lambda$CDM contribution from $T_m^{(0)} \neq T_\text{CMB}$, reads in the limit where $J \gg 1$
\begin{alignat}{1}
    y \approx \int_0^t dt \, \left(\frac{\dot{T}_m^\text{inj}}{H J} - \frac{T_\text{CMB}}{J}\right) \frac{\sigma_T n_e}{m_e} \,,
\end{alignat}
an expression that we use for $J > 100$ to avoid numerical difficulties associated with $T_m$ being close to $T_\text{CMB}$. 
Here, $\Lambda$CDM refers to the standard cosmological model, within which dark matter neither decays nor annihilates.
We note that depending on the size of the energy injection, the $y$-distortion can take on negative values, as is expected in $\Lambda$CDM cosmology, since the matter temperature is always slightly colder than the CMB at early times, see e.g. Eq.~\eqref{eqn:analytic_Tm}. 
For $J \leq 100$ and after recombination, we use the more general Eq.~\eqref{eqn:y-distortion}, with $T_m$ solved using the usual machinery of \dhis to compute the $y$-distortion. 

\subsection{Validation}
\label{sec:y-validation}
\begin{figure}
	\includegraphics[width=\columnwidth]{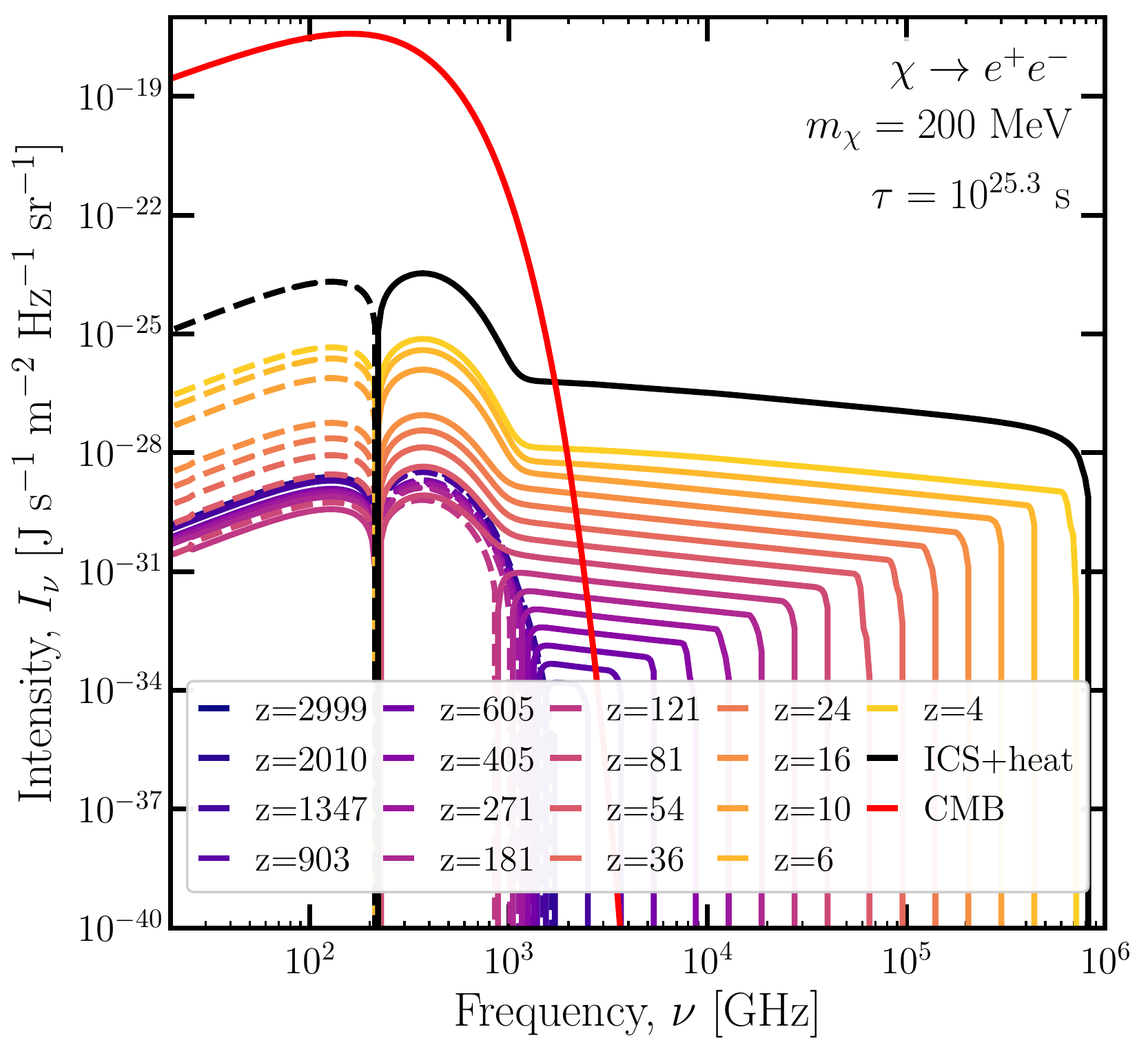}
	\caption{
		Contributions to the spectral distortion from ICS and heating at various redshifts as well as the total distortion.
		We also show the CMB spectrum in red for comparison.
		For visual clarity, we only show every 20th redshift step used to generate this spectral distortion and each contribution has been redshifted to $z=0$.
		The high energy cutoff of each contribution is due to photons with $\omega > \mathcal{R}$ getting absorbed through ionization.
	}
	\label{fig:redshift_dists}
\end{figure}

The spectral distortion from ICS and heating is calculated by first taking the low-energy photon spectra that are output at each redshift step, which we denote by, $J_\text{init} (\omega)$, and then adding in the component from heating using Eq.~\eqref{eqn:y-distortion}; we call their sum $J_\text{new} (\omega)$.
The total distortion from these cooling processes is then given by redshifting the contributions from each step to $z=0$ and summing up the resulting spectra.
Fig.~\ref{fig:redshift_dists} demonstrates this with the example of dark matter with a mass of 200 MeV and a lifetime of $10^{25.3}$ s decaying to $e^+ e^-$ pairs; the colorful lines ranging from blue to yellow show the contribution to the spectral distortion from each redshift step, with each spectrum redshifted to $z=0$.
The dominant component is the $y$-type distortion from the difference in the matter and radiation temperatures, which changes sign around $1+z=155$ when the matter and radiation temperatures decouple.
The black line then shows the sum of all these contributions.
The photons that are being upscattered by the high-energy electron decay products come from the CMB; hence, the troughs of the distortions are located near the same energy as the peak of the CMB blackbody spectrum.
In addition, a photon with initital energy $\omega_0$ will on average upscatter to an energy of $\langle \omega \rangle = \frac{4}{3} \gamma^2 \omega_0$, where $\gamma$ is the Lorentz boost factor for the relativistic electron; hence, the peak of the spectral distortion resulting from ICS is located at $\frac{4}{3} \gamma^2$ times the average energy of photons in the trough~\cite{Blumenthal:1970gc}.

We have also compared the spectral distortions calculated using our method to the Green's functions in Ref.~\cite{Acharya:2018iwh}.
A key difference between our work and Ref.~\cite{Acharya:2018iwh} is that they assume the ionization level is 100\%, so that there is no possibility for photoionization or excitation; at most of the early redshifts they are concerned with, this is a good assumption. 
However, at the relatively later redshifts that we are interested in, there is a small amount of neutral hydrogen prior to recombination, and secondary electrons resulting from the ionization of this hydrogen can deposit a non-negligible amount of energy into heating.
Hence, we find that our spectral distortions are generally larger in amplitude than the Green's functions from Ref.~\cite{Acharya:2018iwh}, see Appendix~\ref{app:cross_checks} for more details.

\section{An Improved Treatment of Low-Energy Photons}
\label{sec:evolution}

The final piece we must consider is that of emission and absorption of photons from atomic transitions.
Our goal in this section is to derive and numerically solve the full set of equations that govern the evolution of injected low-energy photons, and their interactions with atoms, as a function of time, and thereby determine their contribution to the CMB spectral distortion.
By ``injected" photons, we not only mean those directly injected through processes such as dark matter annihilating and decaying to photons, but also those produced as secondary particles from the cooling of high-energy particles, e.g. the spectra described in Sections~\ref{sec:ICS} and \ref{sec:y-type}.
These equations self-consistently solve for the evolution of the low-energy photons and atoms as they interact with one another.
\footnote{In principle, at redshifts low enough for molecules to form, interactions between photons and molecules could also become important; however, we defer this topic to future work.}

\subsection{Summary of previous work and new features}
\label{sec:past_work}

Our treatment here extends that of our previous work~\cite{DH}, which used many convenient approximations that are no longer sufficient when we seek to accurately predict the full spectrum of low-energy photons. Here, we discuss the previous approach and briefly highlight the differences, which we will discuss in more detail in the upcoming sections.

In both our previous work and our current treatment, it is true that at each redshift, exotic energy injection can give rise to a spectrum of high-energy photons, either through direct injection of photons or by the cooling of injected electrons.
We employ tables of precomputed transfer functions to determine how these high-energy photons deposit their energy through cooling processes or into low-energy electrons and photons, as well as how many photons propagate to the next redshift step without cooling.

In our previous work, the resulting low-energy photons are then deposited into different channels, depending on their energy.
We track the energy deposited by different processes using the functions $f_c$, defined by
\begin{equation}
	\left( \frac{dE}{dV dt} \right)_c \equiv f_c \left( \frac{dE}{dV dt} \right)_\text{inj},
\end{equation}
where $\left( \frac{dE}{dV dt} \right)_\text{inj}$ is the energy injected per volume per time and $\left( \frac{dE}{dV dt} \right)_c$ is the energy deposited into channel $c$ per volume per time.

Starting at the lowest energies, previously we assumed that non-CMB photons with energy below $E_\alpha$ redshift until today without further interactions, tracking the energy density of these photons with the function $f_\text{cont}$. 
In contrast, with the new treatment described in this work we can account for interactions these photons have with atoms and the resulting perturbations to their spectrum; 
for example, when a low-energy CMB photon of energy $E$ and an injected photon of energy $E_\alpha-E$ are both absorbed by an atom, exciting it to the $2s$ state.

At the next highest energy, we previously assumed that all photons with energy between $E_\alpha$ and $\mathcal{R}$ are instantly absorbed in exciting neutral hydrogen atoms; this absorbed energy was tracked in the function $f_\text{exc}$. 
These excited atoms can absorb a CMB photon and ionize, leading to an ionization rate of:
\begin{alignat}{1}
\frac1{n_\text{H}} \left( \frac{dE}{dVdt}\right)_\text{inj} (1-\mathcal{C}) \frac{f_\text{exc}}{E_\alpha},
\label{eq:exc_to_ion}
\end{alignat}
where $\mathcal{C}$ is the Peebles-$\mathcal{C}$ factor~\cite{Peebles68} and $\left( \frac{dE}{dVdt}\right)_\text{inj}$ is the exotic energy injection rate per unit volume and time. 
This treatment was also approximate, and did not allow us to keep track of the addition of photons through de-excitation of atoms, nor the subtraction of photons through photo-excitation. 
In the current work we expand the calculation to allow us to keep track of the detailed spectrum of photons absorbed and emitted through such excitations and de-excitations, and also refine the ionization rate of Eq.~\eqref{eq:exc_to_ion}.

Finally, low-energy photons with energies above $\mathcal{R}$ were assumed to all ionize hydrogen atoms (or helium atoms, depending on options set within \texttt{DarkHistory}) and their energy was tracked through the function $f_\text{H ion}$. 
This contributes to the ionization rate through the term:
\begin{alignat}{1}
	\frac{f_\text{H ion}}{\mathcal{R} n_\text{H}} \left( \frac{dE}{dVdt}\right)_\text{inj} .
	\label{eq:just_ion}
\end{alignat}

Our previous approach, based on the simple energy accounting described above, treated hydrogen atoms using the 
Three-Level Atom (TLA) model~\cite{Peebles68, Zeldovich:1969en}. 
This TLA-based treatment assumed that the background radiation bath was a blackbody, and that all excited states above $n=2$ were in perfect thermal equilibrium. 
In our current work, consistent with tracking the detailed interactions of low-energy photons with the gas, we account for the effect that a non-thermal background of photons and a non-equilibrium population of excited states have on the recombination and ionization rates. 

Specifically, to keep track of the higher-$n$ abundances, we drop the TLA approximation and generalize to the Multi-Level Atom (MLA) 
treatment~\cite{Seager:1999km, Chluba:2006bc, Grin_2010, ChlubaDursi_2010}. 
We extend the MLA in two ways. 
First, we allow the phase space density of photons to be non-thermal, allowing changes in the number density of photons to alter the atomic transition rates.
Second, we add source terms due to excitations and ionizations caused by non-thermal injected particles.
In Appendix~\ref{app:cross_checks}, we show that at redshifts where the TLA assumptions hold, our extended MLA treatment and the earlier TLA method yield the same results.

\subsection{Notation and conventions}
\label{sec:dist_exc_intro}

Here, we define the CMB spectral distortion and describe our notation for atomic states and transition rates.
Since we use natural units, we will hereafter denote photon energy by $\omega$.
Denote the full photon phase space density as $f^\gamma(\omega, t)$, and assume that it is homogeneous and isotropic. 
We assume that inhomogeneities at early times are negligible, and ignore any inhomogeneities introduced at late times due to structure formation. 
It is convenient to separate out the dominant blackbody contribution from the rest of the phase space density, 
\begin{alignat}{1}
f^\gamma(\omega, t) = f^\text{CMB}(\omega,t) + \Delta f(\omega,t) \, ,
\label{eq:f_gamma} 
\end{alignat}
where $f^\text{CMB}(\omega, t) = \left[ \exp(\omega/T_\text{CMB}(t))-1 \right]^{-1}$ and $\Delta f(\omega,t)$ is the deviation from the blackbody distribution. 
From now on, we will refer to $\Delta f(\omega,t)$ as the CMB distortion (although note that it may have support at energies far from the peak of the CMB spectrum).
We also define the photon spectrum
\begin{alignat}{1}
N_\omega = \frac{g_\gamma \omega^2}{2 \pi^2 n_\text{B}} f^\gamma(\omega, t) \, ,
\label{eq:dNdE_to_f}
\end{alignat}
which is the number of photons per baryon per energy $\omega$, where $n_\text{B}$ is the baryon density and $g_\gamma=2$ is the number of degrees of freedom for the photon. 
One can check the above set of definitions by ensuring that the number density of photons is equal to
$g_\gamma \int \frac{d^3p}{(2\pi)^3} f^\gamma = n_\text{B} \int d\omega \, N_\omega$. 
The photon spectrum is also related to the spectral radiance by
\begin{equation}
N_\omega = \frac{4\pi}{n_b \omega} I_\omega .
\end{equation}

We will denote bound states of the hydrogen atom either by a subscript $i$ for generic states or subscript $nl$ when the quantum numbers are specified. 
We do not need to consider the magnetic quantum number $m$ in any of our calculations since any splitting effects would produce negligibly small effects at our working precision. 
The existence of $m$ substates is taken into account by multiplicity factors for each $nl$ state.

We denote the radiative transition rate from state $i$ to state $j$ as $R_{i \to j}$, the radiative recombination coefficient to state $i$ as $\alpha_i$, and the  photoionization rate from state $i$ as $\beta_i$; these quantities depend on the photon spectrum and $T_m$.
The energy difference corresponding to $R_{i \to j}$ is labeled as $\omega_{i \to j}$.
We have defined $R_{i \to j}$ to be the rate of transitions per second per hydrogen atom as in Ref.~\cite{preHyrec} without normalizing per volume as in Refs.~\cite{Wong:2005yr, Jens2006}, and we define $R_{i \to i} = 0$.
When calculating $R_{i \to j}$, we ignore numerous percent-level contributions that are included in precise recombination 
codes~\cite{Seager_1999, Seager:1999km, Chluba:2006bc, Chluba_2009, Chluba_2010, Hyrec}; the net effect of these is small, since we are able to reproduce the atomic lines from recombination at percent-level (see Section~\ref{sec:atom-cross-checks} for details).
For example, we ignore any collisional transitions from thermal electrons, focusing on only radiative transitions. 
While we include all dipole transitions and the single quadrupole transition $2s \to 1s$, we ignore all other quadrupole or higher-order transitions. 
We treat all resonances as perfect lines, ignoring any line broadening or other radiative transfer effects.
In Appendix~\ref{app:atomic_rates} we show how to calculate these rates given an arbitrary $f^\gamma(\omega)$ using the methods of Ref.~\cite{preHyrec}.

\subsection{General Evolution Equations}

The most general form of the equations that govern the population of each hydrogen atomic state $i = $ 1s, 2s, 2p, $\cdots$ up to states with principal quantum number $n = n_{\max}$ can be written as~\cite{Hyrec}
\begin{alignat}{2}
    \dot{x}_{1s} &=&& \sum_{k > 1s}^{n_{\max}} (x_k R_{k \to 1s} - x_{1s} R_{1s \to k}) \nonumber \\
    & && \qquad + x_e^2 n_\text{H} \alpha_{1s} - x_{1s} \beta_{1s} - \dot{x}_\text{inj}^{\text{ion}} \,, \nonumber \\
    \dot{x}_{l > 1s} &=&& \sum_{j \neq l}^{n_{\max}} (x_j R_{j \to l} - x_l R_{l \to j}) + x_e^2 n_\text{H} \alpha_l - x_l \beta_l \,,
    \label{eq:general_xi_evolution}
\end{alignat}
where $x_i \equiv n_i / n_\text{H}$ is the ratio of the number density of hydrogen atoms in state $i$ to $n_\text{H}$; all summations over hydrogen states in this paper are taken up to $n = n_{\max}$, and we will suppress indicating this maximum value in the summation for ease of notation. 
We will also reserve $i$, $j$ as indices to represent all states, while $k$ and $l$ represent only excited states. 
$\dot{x}_{\text{inj}}^{\text{ion}}$ is the net contribution to hydrogen ionization from exotic energy injection, including collisional ionization from secondary electrons, and photoionization from photons with $\omega \geq \mathcal{R}$; for $\omega < \mathcal{R}$, we include the effect of injected photons in $\alpha_i$, $\beta_i$ and $R_{i \to j}$, as we will detail below. Note that we assume that all exotic photoionization events occur with a hydrogen atom in the ground state.
In principle, we could attribute some of these ionizations to excited states, which would alter the amount of energy deposited into ionization; however, since the ground state is exponentially populated compared to the excited states throughout recombination, we choose to consider only ground state ionizations.
Differentiating the relation $x_e + \sum_i x_i = 1$ with respect to time, we can determine the evolution of the free electron fraction, 
\begin{alignat}{1}
    \dot{x}_e = - x_e^2 n_\text{H} \sum_i \alpha_i + \sum_i x_i \beta_i + \dot{x}_\text{inj}^\text{ion} \,.
    \label{eq:general_xe_evolution}
\end{alignat}
Since all of the atomic bound-bound rates and photoionization coefficients are functions of the photon spectrum, we also have to specify how the photon spectrum with $\omega \leq \mathcal{R}$ evolves. 
Throughout this paper, we only track distortions to the CMB; any subsequent mention of a spectrum of photons should be understood as a distortion (both positive and negative) to the CMB. 
The evolution of the photon spectrum can be written as
\begin{alignat}{1}
    \dot{N}_\omega = - H \omega \frac{d N_\omega}{d \omega} + J(\omega) \,.
    \label{eq:general_photon_spec_evolution}
\end{alignat}
The first term on the right-hand side accounts for the redshifting of the spectrum, while $J(\omega)$ is the net number of photons injected per baryon, per energy, and per unit time. $J(\omega)$ includes the absorption or emission of photons from bound-bound transitions, recombination/photoionization, and exotic injection: 
\begin{alignat}{2}
    J(\omega) &=&& \frac{n_\text{H}}{n_\text{B}} \sum_i \sum_{j < i} (x_i R_{i \to j} - x_j R_{j \to i}) \delta_D(\omega - \omega_{i \to j}) \nonumber \\
    & &&+ \frac{n_\text{H}}{n_\text{B}}\sum_i \left[ x_e^2 n_\text{H} \gamma_i(\omega) - x_i \xi_i(\omega) \right] + J_\text{new}(\omega) \,. 
    \label{eqn:j_omega}
\end{alignat}
For bound-bound transitions in the first term, we take the spectrum of photons emitted/absorbed to be a line with $\delta_D$ being a Dirac-delta function, since the spectrum is much narrower than our binning. Next, every bound-free transition to/from state $i$ which produces or absorbs a photon of energy $\omega$ also produces or removes a free electron with kinetic energy defined as $\kappa^2 \mathcal{R}$, where $\kappa^2$ is a positive real number. The recombination spectrum is therefore proportional to $x_e^2 n_\text{H}$, with 
\begin{alignat}{1}
    \gamma_i(\omega) \equiv \int_0^\infty d \kappa^2 \, \frac{d \alpha_i}{d \kappa^2} \delta_D(\omega - (\kappa^2 + 1 / n_i^2) \mathcal{R}) \, ,
\end{alignat}
where $n_i$ is the principal quantum number of state $i$.
Similarly, the spectrum of photons absorbed in photoionization is proportional to $x_i$, with 
\begin{alignat}{1}
    \xi_i(\omega) \equiv \int_0^\infty d \kappa^2 \, \frac{d \beta_i}{d \kappa^2} \delta_D(\omega - (\kappa^2 + 1 / n_i^2) \mathcal{R}) \,.
\end{alignat}
$J_\text{new}(\omega)$ is the number of photons deposited per baryon, per energy, and per unit time; our upgraded version of \dhis computes precisely this low-energy photon spectrum given a source of injected energy. 
It receives contributions from ICS of injected high-energy electrons off the CMB (see Section~\ref{sec:ICS}) and from $y$-distortions due to heating of the IGM (see Section~\ref{sec:y-type}). 
We note that previous versions of \dhis only explicitly tracked the contribution from ICS, without including the backreaction of these distortions on bound-bound and bound-free transitions. 

To close these equations, we use Eq.~\eqref{eq:general_Tm_evolution} as the evolution equation for the IGM temperature.
Eqs.~\eqref{eq:general_xi_evolution},~\eqref{eq:general_xe_evolution},~\eqref{eq:general_photon_spec_evolution} and~\eqref{eq:general_Tm_evolution} are complete and can be solved to determine the joint evolution of hydrogen atoms and radiation in the presence of exotic energy injection. However, there are vast differences in timescales in the various terms that both make it difficult to solve this general set of equations, and also allow the use of well-known simplifications to reduce the equations into a more tractable form~\cite{Peebles68,preHyrec,Hyrec}.

\subsection{Simplified Evolution Equations}

\subsubsection{Rapid photoionization}

The first approximation arises from the fact that once an appreciable population of bound neutral hydrogen forms, which occurs at around $1+z \sim 1500$, the mean free time of a photoionizing photon is short compared to the Hubble timescale, leading to two effects, both outlined in Ref.~\cite{Peebles68}. 
First, the distribution of photons with $\omega > \mathcal{R}$ is driven strongly toward equilibrium with the IGM, which has a temperature $T_m \ll \mathcal{R}$, implying that all photoionizing photons produced by exotic injections are to a good approximation absorbed. This allows us to write
\begin{alignat}{1}
    \dot{x}_\text{inj}^\text{ion} = \frac{f_\text{H ion}(z, x_e)}{\mathcal{R} n_\text{H}} \left(\frac{dE}{dV \, dt}\right)_\text{inj} \,,
    \label{eqn:x_inj_term}
\end{alignat}
where the right-hand side gives the number of ionizations from exotic energy injections per H nucleus per unit time.
Second, for every recombination to the ground state, an ionizing photon is emitted that is immediately absorbed by another atom in the ground state, leading to no net recombination; this is known as case-B recombination. 
This allows us to drop $- x_e^2 n_\text{H} \alpha_{1s} + x_\text{1s} \beta_{1s}$ in Eq.~\eqref{eq:general_xi_evolution}, which cancel each other down to a residual much smaller than the Hubble rate~\cite{Peebles68}. 
This also simplifies the $x_e$ evolution equation to 
\begin{alignat}{1}
    \dot{x}_e &= - x_e^2 n_\text{H} \sum_{k > 1s} \alpha_k + \sum_{k > 1s} x_k \beta_k + \dot{x}_\text{inj}^\text{ion}, 
    \label{eq:simplified_intermediate_xe}
\end{alignat}
which depends on the properties of the IGM. 
Throughout, we assume that free electrons are distributed thermally with temperature $T_m$.

\subsubsection{Rapid Lyman-series transitions}

Let us turn now to Lyman-series, i.e.\ $1s \to np$ photons. When an appreciable population of ground state hydrogen forms, the mean free path of Lyman series radiation is, as for photoionizing photons, much shorter than the Hubble length. As with photoionizing photons, Lyman-series photons are driven strongly toward equilibrium with the IGM, so that any exotic injections of Lyman-series photons are to a good approximation absorbed rapidly. Furthermore, any $np \to 1s$ transition produces a Lyman-series photon that is quickly reabsorbed, significantly suppressing the effectiveness of $np \to 1s$ processes in depleting the $np$ state. Unlike recombination photons, however, Lyman-series photons are emitted/absorbed in an extremely narrow frequency range, granting emitted photons the opportunity to redshift out of the narrow Lyman-series resonances via the first term in Eq.~\eqref{eq:general_photon_spec_evolution}. The net rate of transitions from $np$ to $1s$ can be properly accounted for by using the Sobolev approximation~\cite{Seager:1999km}, where the net rate is multiplied by the probability of a Lyman-series photon redshifting out of a line, $p_{np} \equiv [1 - \exp(-\tau_{np})] / \tau_{np}$, where $\tau_{np}$ is the Sobolev optical depth for a photon with energy corresponding to the $1s \to np$ transition. 

To take both effects into account in our calculation, we define
\begin{alignat}{1}
    \tilde{R}_{i \to j} \equiv \begin{cases}
        p_{np} R_{i \to j} \,, & i = 1s, j = np \text{ or } j=1s, i=np \,, \\
        R_{i \to j} \,, & \text{otherwise} \,,
    \end{cases}
\end{alignat}
and replace $R_{i \to j} \to \tilde{R}_{i \to j}$ in Eq.~\eqref{eq:general_xi_evolution}. We also assume that all Lyman-series photons $J_{np}$ from exotic injection are rapidly absorbed, leading directly to $1s \to np$ excitations. These photons are therefore not included in $J(\omega)$, but instead directly contribute to a term $\dot{x}_{\text{inj}, np}^\text{exc}$, where
\begin{alignat}{1}
    \dot{x}_{\text{inj},np}^\text{exc} \equiv \frac{n_\text{B}}{n_\text{H}} J_{np} + \dot{x}_{np}^{\text{coll. exc.}}\,. 
\end{alignat}
The term $\dot{x}_{np}^{\text{coll. exc.}}$ represents the contribution of collisional excitation from low-energy electrons.
Note that we do not include this correction for resonant photons that connect two excited states, since the abundance of excited states is extremely suppressed compared to the abundance of the ground state, and resonant lines between two excited states do not interact strongly enough to drive these photons to equilibrium with the IGM. Further details on this treatment can be found in Appendix~\ref{app:atomic_rates}. 

At this point, including both the approximations for fast photoionization and fast Lyman-series scattering, the evolution equations for $x_i$ read
\begin{alignat}{2}
    \dot{x}_{1s} &=&& \sum_{k > 1s} (x_k \tilde{R}_{k \to 1s} - x_{1s} \tilde{R}_{1s \to k} - \dot{x}_{\text{inj},k}^\text{exc}) - \dot{x}_\text{inj}^\text{ion} \,, \nonumber \\
    \dot{x}_{l>1s} &=&& \sum_{j \neq l} (x_j \tilde{R}_{j \to l} - x_l \tilde{R}_{l \to j}) + x_e^2 n_\text{H} \alpha_l - x_l \beta_l + \dot{x}_{\text{inj},l}^\text{exc}\,, 
    \label{eq:x_i_after_rapid_ion_exc_approx}
\end{alignat}
where we set $\dot{x}_{\text{inj},l \neq np}^\text{exc} \equiv 0$ for ease of notation. 

\subsubsection{Steady state approximation}
\label{sec:steady_state}

The next significant approximation we make is the steady-state approximation for all excited states~\cite{Grin_2010,preHyrec,Hirata:2008ny}. 
The total rate for transitioning out of an excited state $k$ is
\begin{alignat}{1}
    \tilde{R}_{k}^\text{out} \equiv \sum_j \tilde{R}_{k \to j} + \beta_k \,,
\end{alignat}
which is much faster than the Hubble rate at all redshifts~\cite{preHyrec,Hyrec}.
The populations of the excited states in the hydrogen atom are therefore driven toward a fixed point for each $x_k$ set by $\dot{x}_k = 0$; any differences from these fixed points are rapidly erased on timescales much shorter than a Hubble time. This steady state approximation reduces the population of the atomic states to a set of algebraic equations of the form $x_k = M^{-1}_{kl} b_l$, with $M_{kl}$ and $b_l$ are objects indexed by excited states, evolving on Hubble timescales.  

Let us now explicitly write out the matrix $M$ and the inhomogeneous term $b$. Setting $\dot{x}_i = 0$ in Eq.~\eqref{eq:x_i_after_rapid_ion_exc_approx} and moving the negative terms to the right hand side, we obtain for the excited states
\begin{alignat}{1}
    x_k \tilde{R}_k^\text{out} = \sum_{l > 1s} x_l \tilde{R}_{l \to k} + x_{1s} \tilde{R}_{1s \to k} + x_e^2 n_\text{H} \alpha_k + \dot{x}_{\text{inj},k}^\text{exc} \,,
\end{alignat}
which simply equates the rate of leaving the state $i$ to the rate of all transitions into the state $i$. This is a linear system in excited states $x_k$, which we can now invert. To make the relation between this expression and previous results clear, however, we write the solution as
\begin{alignat}{1}
    x_k = \sum_{l > 1s} M^{-1}_{kl} (b_l^{1s} + b_l^\text{rec} + b_l^\text{inj}) \,,
    \label{eq:matrix_MLA}
\end{alignat}
where 
\begin{alignat}{1}
    M_{kl} &= \delta_{kl} \tilde{R}_k^\text{out} - \tilde{R}_{l \to k} \,, \nonumber \\
    b_l^{1s} = x_{1s} \tilde{R}_{1s \to l} \,, &\quad b_l^\text{rec} = x_e^2 n_\text{H} \alpha_l \,, \quad b_l^\text{inj} = \dot{x}_{\text{inj},l}^\text{exc} \,.
    \label{eq:M_and_b}
\end{alignat}
where $\delta_{kl}$ is the Kronecker delta function. The three source terms populate the excited state $l$ in three different ways: $b_l^{1s}$ via excitations from the ground state sourced by $f^\gamma (\omega,t)$, $b_l^\text{rec}$ via recombinations, and $b_l^\text{inj}$ via interactions with injected particles, including photoexcitations by Lyman-series photons and collisional excitations by injected electrons. 

With the aid of the identity
\begin{alignat}{1}
    \beta_k = \tilde{R}_k^\text{out} - \tilde{R}_{k \to 1s} - \sum_{l > 1s} \tilde{R}_{k \to l} = \sum_l M_{lk} - \tilde{R}_{k \to 1s} \,,
\end{alignat}
we can now also simplify Eq~\eqref{eq:simplified_intermediate_xe} to obtain 
\begin{alignat}{1}
    \dot{x}_e = - x_e^2 n_\text{H} \tilde{\alpha}_\text{B} + x_{1s} \tilde{\beta}_\text{B} + \dot{x}_\text{inj} \,,
    \label{eq:final_xe}
\end{alignat}
where
\begin{alignat}{1}
    \tilde{\alpha}_\text{B} &= \sum_{k > 1s}  Q_k \alpha_k \,, \nonumber \\
    \tilde{\beta}_\text{B} &= \sum_{k > 1s} (1 - Q_k) \tilde{R}_{1s \to k} \,, \nonumber \\
    \dot{x}_\text{inj} &= \sum_{k > 1s} (1 - Q_k) \dot{x}_{\text{inj}, k}^\text{exc} + \dot{x}_\text{inj}^\text{ion} \,, \nonumber \\
    Q_k &= \sum_{l} M_{lk}^{-1} \tilde{R}_{l \to 1s} \,. 
\end{alignat}
$Q_k$ is a weight that determines how effective recombination to the ground state is from state $k$; if $Q_k = 1$ for all $k$, then recombination to state $k$ ultimately always results in a ground state. 
$Q_k$ is related to the Peebles-$C$ factor in the three-level atom model. 
Moreover, we use the suggestive notation $\tilde{\alpha}_\text{B}$ and $\tilde{\beta}_\text{B}$ because under the assumptions of the three-level atom, these quantities are closely related to the case-B recombination and photoionization rates.

Note that since $Q_k$ is very close to the identity, the quantity $1-Q_k$ can be very slow to compute.
In Appendix~\ref{app:matrix_method}, we outline an alternative procedure for calculating these quantities that is numerically faster and more in line with what is done in the code.
Further discussion of these equations and how they reduce to the three-level atom model can be found in Appendix~\ref{app:derivation_TLA}. 

To close these equations, we use the approximation $x_{1s} \approx 1 - x_e$, which is an excellent approximation, since we find that the excited states are typically populated at the level of $10^{-13}$ of the ground state shortly after recombination (See e.g. Fig.~\ref{fig:MLA_vs_TLA} in Appendix~\ref{app:cross_checks}). 
With this approximation, the only quantity we have to track explicitly is $x_e$, although given $x_e$ and the photon spectrum, the population of all hydrogen states can be computed. We stress that the effect of exotic particle injection on the ionization term is not solely contained in $\dot{x}_\text{inj}$, but also manifests itself in the fact that low-energy photons arising from the injection affect $\tilde{\alpha}_\text{B}$ and $\tilde{\beta}_\text{B}$, which both depend on the photon spectrum. Altogether, the equations we finally solve are
\begin{alignat}{1}
    \dot{x}_e &= - x_e^2 n_\text{H} \tilde{\alpha}_\text{B} + (1 - x_e) \tilde{\beta}_\text{B} + \dot{x}_\text{inj} + \dot{x}^\text{re} \,, \nonumber \\
    \dot{T}_m &= - 2 H T_m + \Gamma_C(T_\text{CMB} - T_m) + \dot{T}_m^\text{inj} + \dot{T}_m^\text{re} \,, \nonumber \\
    \dot{N}_\omega &= - H \omega \frac{d N_\omega}{d\omega} + J(\omega) \,.
    \label{eq:full_eqs}
\end{alignat}
We have now included $\dot{x}^\text{re}$ and $\dot{T}_m^\text{re}$ to account for ionization and heating from reionization sources at late times; the definition of these terms is unchanged relative to \dhis\texttt{v1.0}~\cite{DH}.
\footnote{In principle, one should also modify the Multi-Level Atom (MLA) treatment of the excited hydrogen states to include the radiation fields that cause reionization in the first place. Since we have not tested our numerical method described in Section~\ref{sec:numerical MLA} with these radiation fields, we leave including this source to future work.} 

Note that here we have omitted the helium ionization equations; they have also not changed from our previous treatment~\cite{DH}. While atomic transitions between states of helium could in principle also contribute to the spectral distortion, energy deposition into helium ionization and excitation is subdominant to the hydrogen contribution, so we expect the effect on the photon spectrum to be small, see e.g. Ref.~\cite{Rubino-Martin:2007tua}. To briefly summarize, the helium ionization equation is a sum of three contributions: 1) the expected contribution in the absence of energy injection, which is identical to the \texttt{Recfast} treatment~\cite{Seager_1999}, 2) a source term from processes that are active at reionization, and 3) a term from exotic energy injection which is analogous to the hydrogen ionization term given in Eq.\eqref{eqn:x_inj_term} but with $f_\text{H ion} \rightarrow f_\text{He ion}$, where $f_\text{He ion}$ is the fraction of injected energy that is deposited into helium ionization, and $\mathcal{R} \rightarrow \mathcal{R}_\text{He} = 24.6$ eV.
\dhis includes a few possible methods to treat the contributions of low-energy photons to $f_\text{He ion}$ that bracket the uncertainties on how helium ionization proceeds.
The default treatment is to ascribe the energy from all photoionizations to hydrogen, so the only helium ionizations occur through collisional ionizations by electrons.
While this may not be the most accurate treatment, it is the simplest and is valid well before reionization.

\subsection{Numerical Method}
\label{sec:numerical MLA}

%
\begin{figure*}
	\includegraphics[width=\textwidth]{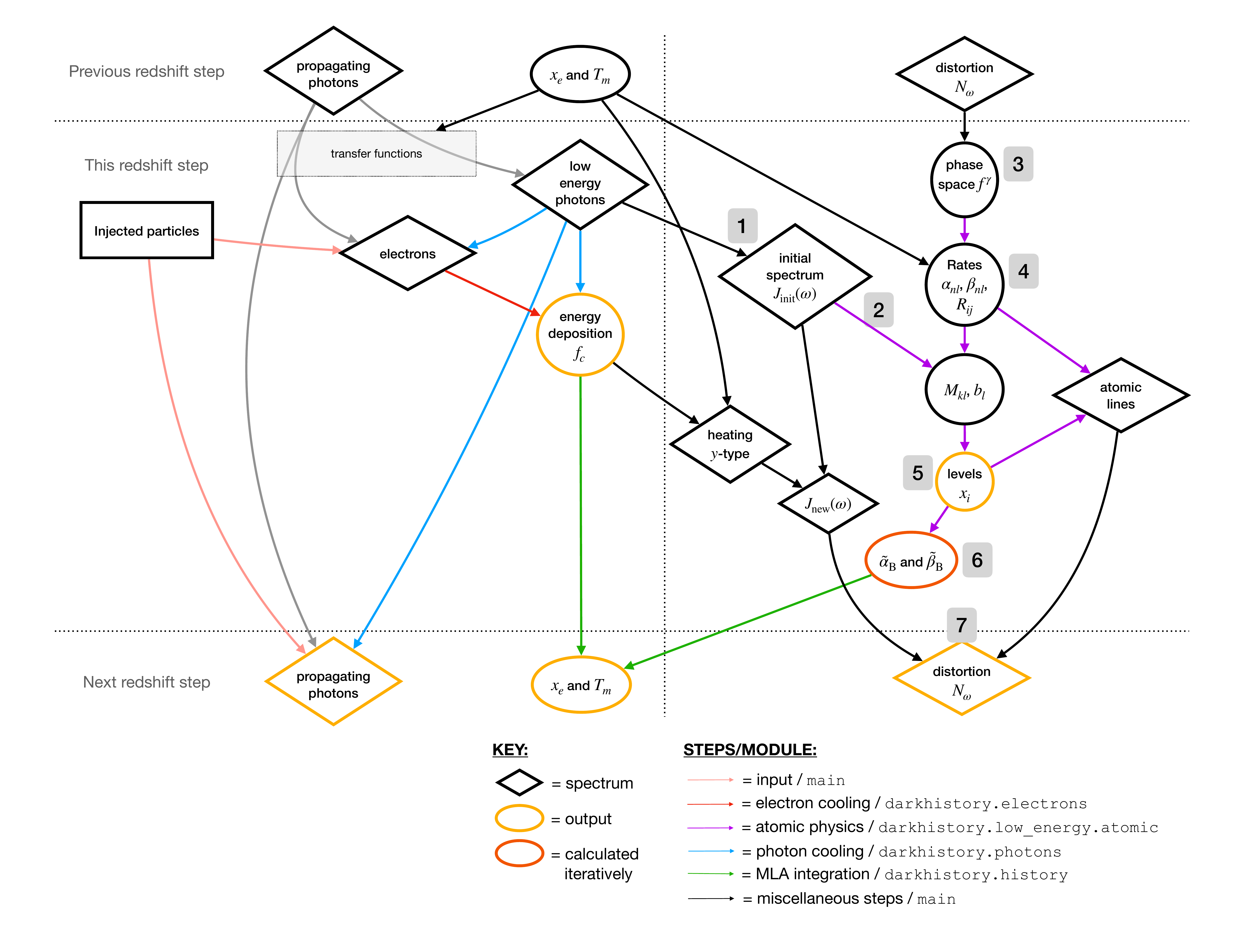}
	\caption{
		Flowchart summarizing our numerical procedure. 
		The outputs of \dhis are highlighted in the yellow shapes.
		We also highlight $\tilde{\alpha}_\text{B}$ and $\tilde{\beta}_\text{B}$ in orange-red to indicate that this quantity is calculated iteratively, as described at the end of Section~\ref{sec:numerical MLA}.
		The right side highlights the new quantities and atomic processes that we track in \texttt{DarkHistory}, as described in Section~\ref{sec:evolution}.
		The numbers label the steps outlined in Section~\ref{sec:numerical MLA}.
	}
	\label{fig:flowchart}
\end{figure*}

We now describe our numerical procedure for integrating the system of differential equations in Eq.~\eqref{eq:full_eqs}, and focus in particular on how we obtain the spectral distortion.
This procedure is implemented in \texttt{main.evolve()} provided the \texttt{distort} option is set to \lstinline|True|.

Initially, the photon phase space density is set to its blackbody value, so $\Delta f(\omega,t) = 0$; the ionization level is set to its Saha equilibrium value; and the initial matter temperature is determined using the early time analytic formula, Eq.~\eqref{eqn:analytic_Tm}.
The initial redshift for the integration should be early enough to justify using the above initial conditions; by default, we use $1+z_\text{init} = 3000$.

To integrate from the initial redshift to today, we use a fixed step size in log redshift space, which by default is set to $\Delta \ln(1+z) = .001$. 
At each redshift step, we perform the following calculations.
\begin{enumerate}
	\item Calculate the initial spectrum of injected low-energy photons, $J_\text{init} (\omega)$.
	Note that this does not include directly injected photons, but rather photons that are deposited as result of the cooling of directly injected particles.
	\item Zero out the ionizing bins in the above spectrum with bin centers above $\mathcal{R}$ and the Lyman-series bins that contain energies $\mathcal{R} ( 1 - n^{-2})$, for $n \in \{1, 2,...,n_\text{max}\}$, accounting for the energy in these bins by modifying $f_\text{H ion}$ and the $b_l^\text{inj}$ term of Eq.~\eqref{eq:matrix_MLA}.
	\item Use the full spectrum of low-energy photons, $N_\omega$, to calculate the photon phase space density using Eq.~\eqref{eq:dNdE_to_f}.
	This spectrum includes the total spectral distortion accumulated up to this redshift.
	\item Compute the transition rates $\tilde{R}_{i \to j}$, $\alpha_i$, and $\beta_i$ using the equations within Appendix~\ref{app:atomic_rates} in the presence of a non-zero $\Delta f(\omega,t)$.
	\item Set $x_{1s} = 1-x_e$, solve for $M$ and $b$ using Eq.~\eqref{eq:M_and_b}, then determine the population levels $x_i(z)$ by inverting Eq.~\eqref{eq:matrix_MLA}.
	\item Use $M$, $\tilde{R}_{i \to j}$, $\alpha_i$, and $\beta_i$ to compute $\tilde{\alpha}_\text{B}$, $\tilde{\beta}_\text{B}$, and $\dot{x}_\text{inj}$,
	then solve for $x_e$ and $T_m$ at the end of the redshift step by integrating the first two equations of Eq.~\eqref{eq:full_eqs} through the redshift step,
	\item Using the occupation numbers $x_i(z)$ and rates already computed, calculate the photons produced from atomic processes using Eq.~\eqref{eqn:j_omega}.
	In addition, use $T_m$ to determine the amplitude of the $y$-type distortion contributed at this redshift step through Eq.~\eqref{eqn:y_DM}; this together with $J_\text{init} (\omega)$ constitutes $J_\text{new} (\omega)$.
	Add the atomic contribution and $J_\text{new} (\omega)$ to $N_\omega$.
	\item Redshift all spectra to the next redshift step.
\end{enumerate}
Looping through these steps evolves all quantities forward in time, gradually building up the spectral distortion $N_\omega$ until we reach the chosen ending redshift; in our analysis, we choose $1+z_\text{end}=4$ since our treatment of ionized helium breaks down past this point.
We then redshift the photon spectrum to today. 
The outputs of this version of the \dhis code are $T_m$ and $x_e$ as a function of time and the present day photon spectrum.
This procedure is summarized in Fig.~\ref{fig:flowchart}.

\begin{figure}
	\includegraphics[width=\columnwidth]{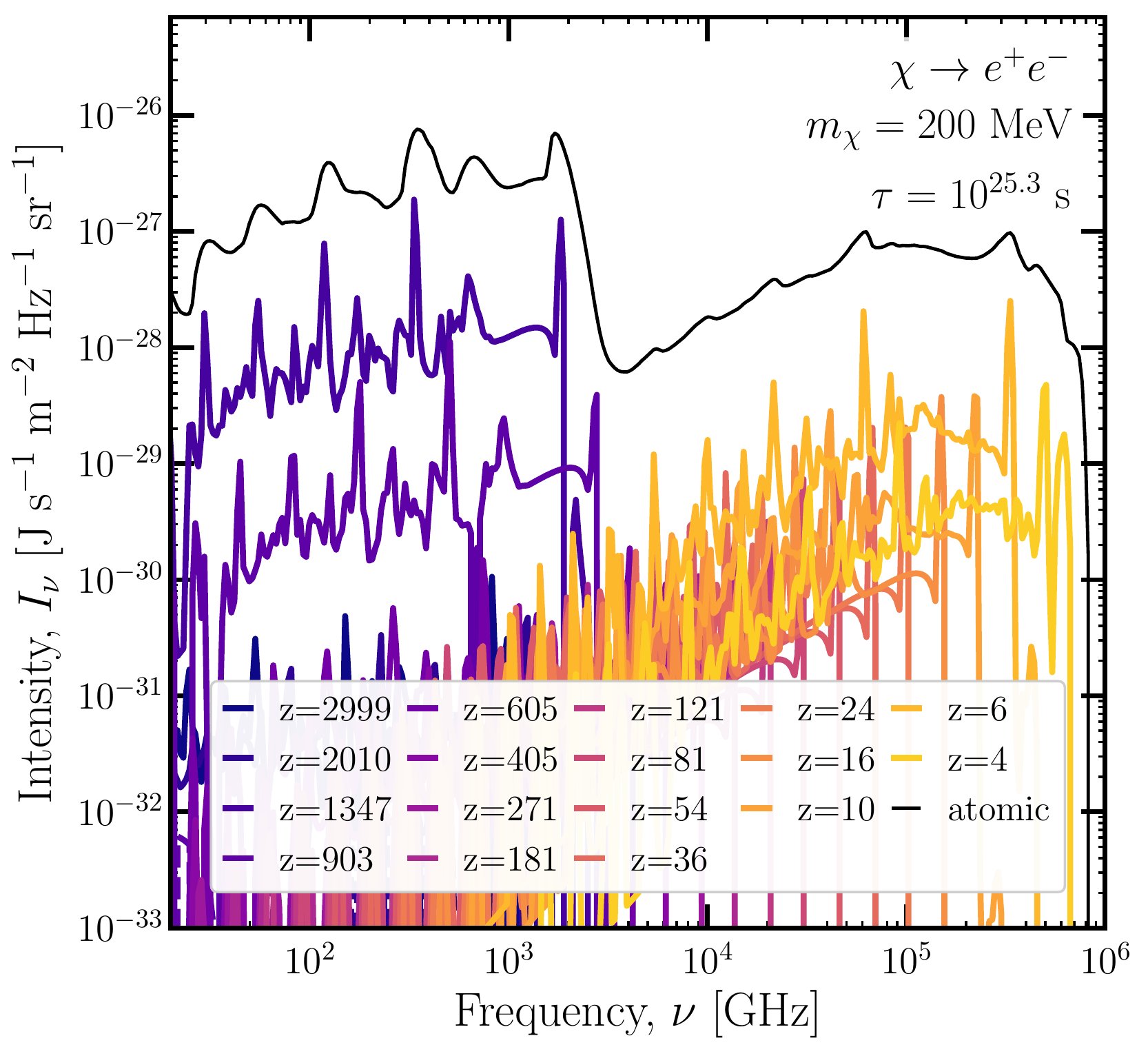}
	\caption{
		Changes to the spectral distortion from atomic transitions at each redshift step, as well as the total distortion.
		For visual clarity, we only show every 20th redshift step used to generate this spectral distortion and each contribution has been redshifted to $z=0$.
	}
	\label{fig:redshift_dists_atomic}
\end{figure}
\begin{figure}
	\includegraphics[width=\columnwidth]{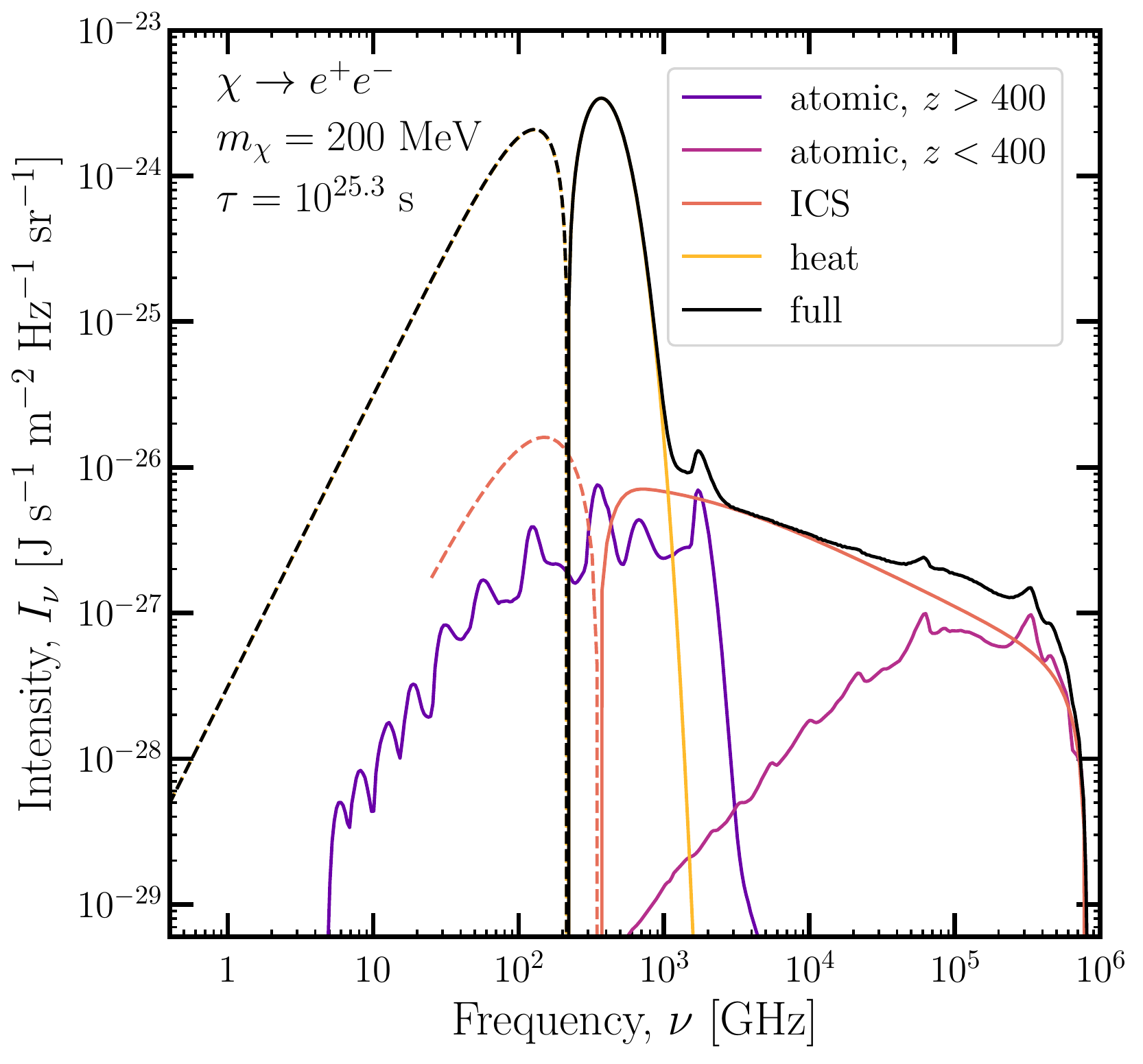}
	\caption{
		The different contributions to the spectral distortion; here we choose dark matter decaying to $e^+ e^-$ pairs, with a mass of $200$ MeV and a lifetime of $10^{25.3}$ s.
		The contributions we show are atomic lines from $z > 400$ which are dominated by the redshifts around recombination, the atomic lines from $z < 400$ which are dominated by the redshifts around reionization, the photons resulting from ICS, and the $y$-type distortion resulting from heating of the IGM.
		The sum of the ICS and heat contributions is the same as the black line in Fig.~\ref{fig:redshift_dists}.
	}
	\label{fig:components}
\end{figure}

In Fig.~\ref{fig:redshift_dists_atomic}, we show the change in the distortion resulting from atomic transitions at each redshift step, including the effect of the full distortion on the atomic states and their recombination/photoionization coefficients.
We also show the final distortion contributed by atomic lines in black.
All spectra shown are redshifted to $z=0$.
The dark matter model used is the same as that in Fig.~\ref{fig:redshift_dists}.

The total spectral distortion for the energy injection model used in Fig.~\ref{fig:redshift_dists_atomic} is shown in Fig.~\ref{fig:components} in black; we also break down the spectral distortion by the photon sources, including ICS, heating, atomic lines from before $z = 400$ (which are mostly from recombination), and atomic lines from after $z=400$ (mostly from reionization).
The component with the largest amplitude is generated by heating.
We note that many of these photons are generated by $\Lambda$CDM processes; in other words, some of this spectral distortion would still be present if we turned off exotic energy injections.

Let us pause to analyze how this procedure self-consistently captures the effect of the radiation on the evolution of the atoms, as well as the atoms' effect on the radiation.  
To capture the effect of radiation on the atoms, we add $\Delta f(\omega,t)$ to the black-body phase space density in our calculation of $\tilde{R}_{i \to j}$, altering the rates of de-excitation, excitation, recombination, and ionization. 
Since $\Delta f(\omega,t)$ can be negative or positive, these rates can be either diminished or enhanced.
The changes in these rates subsequently modifies $M_{kl}$ and the $b_l$ terms, and hence modifies the populations of atomic states and ionization level, $x_i$ and $x_e$.
Beyond the effects on the photon spectrum, we include exotic energy injections in the evolution equations via the term $\dot{x}_\text{inj}$, which captures ionization and excitation from the additional energy injection.

The atoms can in turn affect the radiation if the rates and populations are such that there is net absorption or emission between two atomic states.
This includes the absorption of photons from previous time steps which redshift into resonant lines.
The resulting photons or photon deficit are then added to $\Delta f(\omega,t)$ within the redshift step.

\begin{figure*}
	\includegraphics[width=\textwidth]{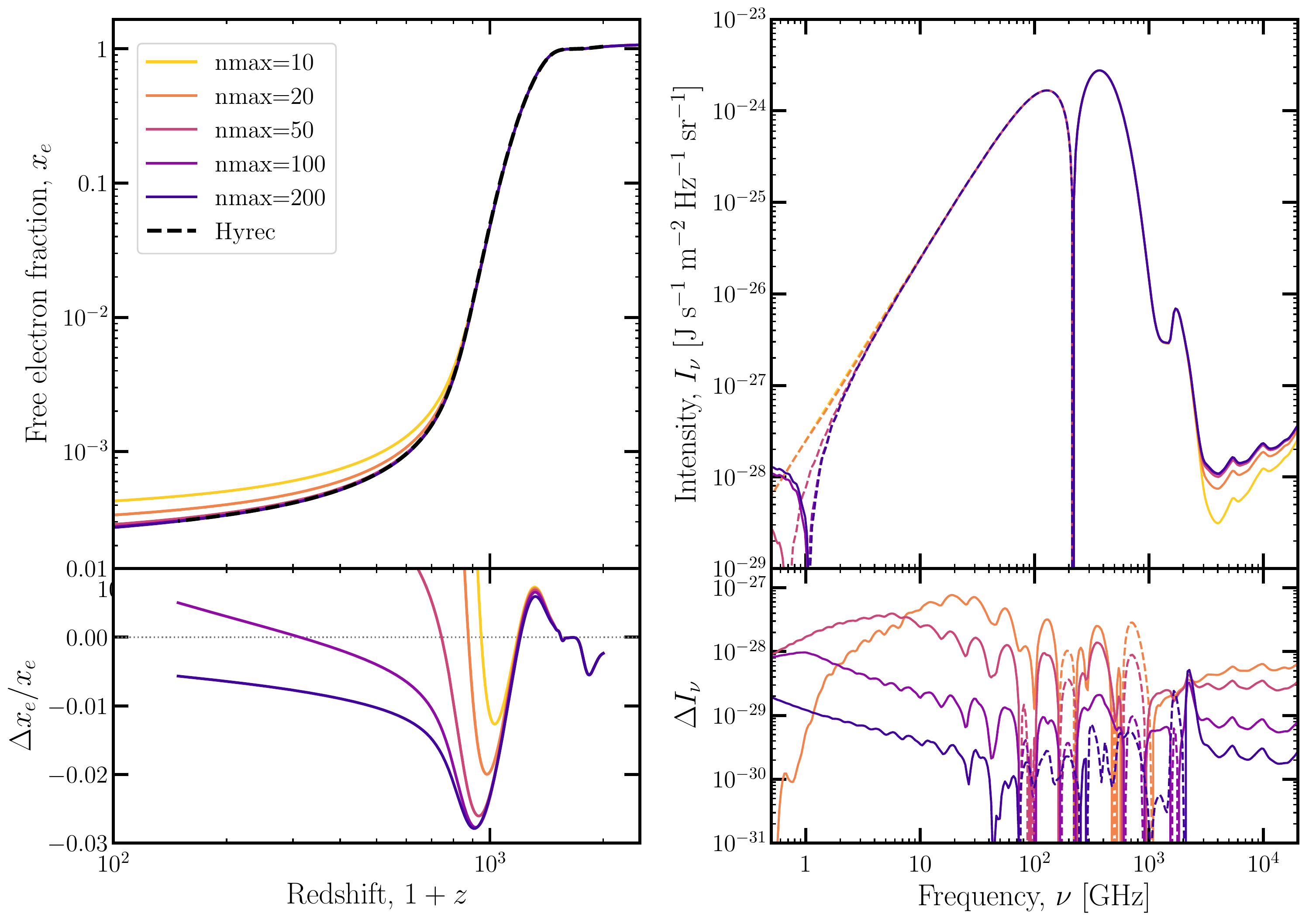}
	\caption{\textit{Left panel:} Convergence of the ionization history with $n_\text{max}$ calculated using our method, compared against the ionization calculated with \texttt{Hyrec}~\cite{Hyrec} which is shown in the black dashed curve.
	The lower panel shows the percent difference between \dhis at the different values of $n_\text{max}$ and \texttt{Hyrec}. 
	The relative difference between the result with  $n_\text{max}=200$ and \texttt{Hyrec} is largest around recombination, but is still only at the level of a few percent.
	\textit{Right panel:} Convergence of the spectral distortion with $n_\text{max}$.
	The color coding is the same as in the left panel.
	The lower panel shows the change in the spectral distortion between using one value of $n_\text{max}$ and the next smallest value.
	Since this difference decreases as we increase $n_\text{max}$, the spectrum of low-energy photons is essentially converged by $n_\text{max}=100$.}
	\label{fig:nmax_convergence}
\end{figure*}
A number of additional practical points need to be addressed. 
One must choose a highest energy state at which to truncate the sum over excited states, $n_\text{max}$. 
We find that with $n_\text{max}=200$ the ionization level is essentially converged, as was also found in Ref.~\cite{ChlubaDursi_2010}, and with $n_\text{max}=100$ the photon spectrum $N_\omega$ is essentially converged.
In Fig.~\ref{fig:nmax_convergence}, we show the ionization histories at different values of $n_\text{max}$ on the left; the black dashed curve is reproduced from Fig. 3 of Ref.~\cite{Hyrec}.
The lower panel shows the relative difference between the curves calculated with \dhis and the \texttt{Hyrec} result~\cite{Hyrec}.
The right panel depicts the spectral distortion calculated at the same values for $n_\text{max}$; the lower panel shows the change in the spectral distortion as we increase $n_\text{max}$ from one value to the next largest value.
In both panels, we can see that the quantities are converged for $n_\text{max}$ greater than about 100.

Once $n_\text{max} \sim \mathcal{O}(10)$, the computation of $\tilde{R}_{i \to j}$, $\alpha_i$, $\beta_i$, and the matrix inversion in Eq.~\eqref{eq:matrix_MLA} become the most computationally expensive steps. 
To speed up our computations, we precompute any quantity that does not depend on redshift, like dipole operator matrix elements connecting bound-bound and bound-free states. 
In addition, $M_{kl}$ is a sparse matrix, meaning most of its elements are equal to zero.
There exist techniques and code packages for taking advantage of sparse structure to speed up linear algebra operations, so we employ \texttt{scipy.sparse} to more efficiently invert Eq.~\eqref{eq:matrix_MLA}.

Another issue is that the differential equations for $x_e$ and $T_m$ are stiff. 
At sufficiently early times the recombination, ionization, and Compton heating rates are so fast compared to a Hubble time that a numerical solution to the evolution equations 
yields $x_e = x_e^\text{Saha}$ and $T_m = T_\text{CMB}$ plus numerical noise. 
To combat this noise, for redshifts $z > 1555$\footnote{This value is close to the one used in Ref.~\cite{preHyrec}, but this is coincidental.} we simply set $x_e = x_e^\text{Saha}$ and $T_m$ to the analytic expression in Eq.~\eqref{eqn:analytic_Tm}.

\begin{figure}
	\includegraphics[width=\columnwidth]{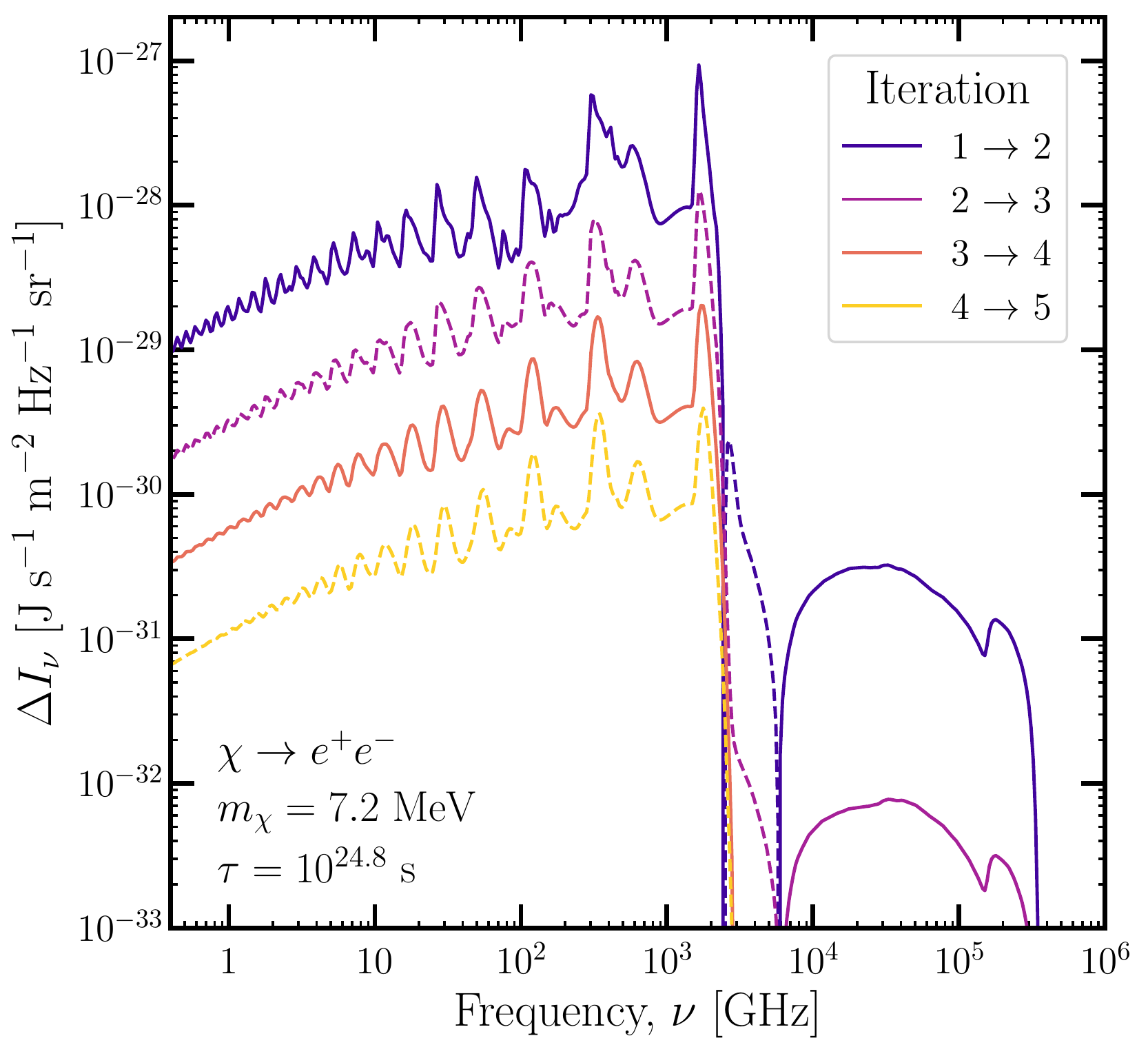}
	\caption{Change in the spectral distortion between one iteration of calculating $\tilde{\alpha}_\text{B}$ and $\tilde{\beta}_\text{B}$ over the full redshift range from $1+z_\text{init}$ to $1+z_\text{end}$ and the next iteration.
	With each iteration, the difference decreases by about an order of magnitude, indicating that the spectral distortion is rapidly converging.}
	\label{fig:iterations}
\end{figure}

Finally, we must be careful with the interpolation and extrapolation of the recombination and ionization rates used in the ionization evolution equation. 
Within each redshift step we calculate $\tilde{\alpha}_B$ and $\tilde{\beta}_\text{B}$. 
Other than Euler's method, which is insufficient for such stiff equations, any integration method requires multiple evaluations throughout the redshift step. 
Each of these evaluations would require recalculating $\tilde{\alpha}_B$ and $\tilde{\beta}_\text{B}$ at slightly different redshifts; however, since each evaluation of $\tilde{\alpha}_\text{B}$ and $\tilde{\beta}_\text{B}$ is so costly, we have devised an iterative method that avoids their computation during the integration loop.
For each step of the iterative method we integrate Eq.~\eqref{eq:full_eqs} over the full redshift range from $1+z_\text{init}$ to $1+z_\text{end}$ using the method described above, but using a precomputed formula functional form for $\tilde{\alpha}_\text{B}$ and $\tilde{\beta}_\text{B}$, instead of computing them during the loop. 
For the first iteration, we use the fitting functions used by \texttt{Recfast}~\cite{Seager_1999}, including the hydrogen fudge factor which we set to 1.125 and the double Gaussian function correction. 
At the end of this iteration, we have a new set of $\tilde{\alpha}_\text{B}$ and $\tilde{\beta}_\text{B}$ values associated with each redshift step that we computed. 
We linearly interpolate these new values with respect to redshift and proceed through the next iteration using these interpolation functions to determine $\tilde{\alpha}_\text{B}$ and $\tilde{\beta}_\text{B}$. 
We find that this process already converges after one iteration (see Fig.~\ref{fig:iterations}). 
%

\subsection{Comparison to Other Calculations}
\label{sec:atom-cross-checks}

In this section we perform the numerical procedure described in the previous section and compare our outputs to those of other codes. 
First we compare our calculation of the ionization history, $x_e(z)$, to the output of the recombination code, \texttt{Hyrec}. 
If we again look at Fig.~\ref{fig:nmax_convergence}, the dark purple line is the ionization history calculated using our code with $n_\text{max} = 200$; the black dashed line shows the same quantity calculated with \texttt{Hyrec}.
The second panel shows the relative difference between the two; the greatest deviations come from around the redshift of recombination, but are still only at the level of a few percent.
These differences may be due to a number of effects we neglected that are accounted for in \texttt{Hyrec}, including helium recombination, two-photon transitions from levels higher than 2s, and frequency diffusion in the Lyman-$\alpha$ line~\cite{Hyrec}.
Hence, \texttt{Hyrec} is the more accurate recombination code, but few-percent-level accuracy is sufficient for our purposes.

\begin{figure}
	\includegraphics[width=\columnwidth]{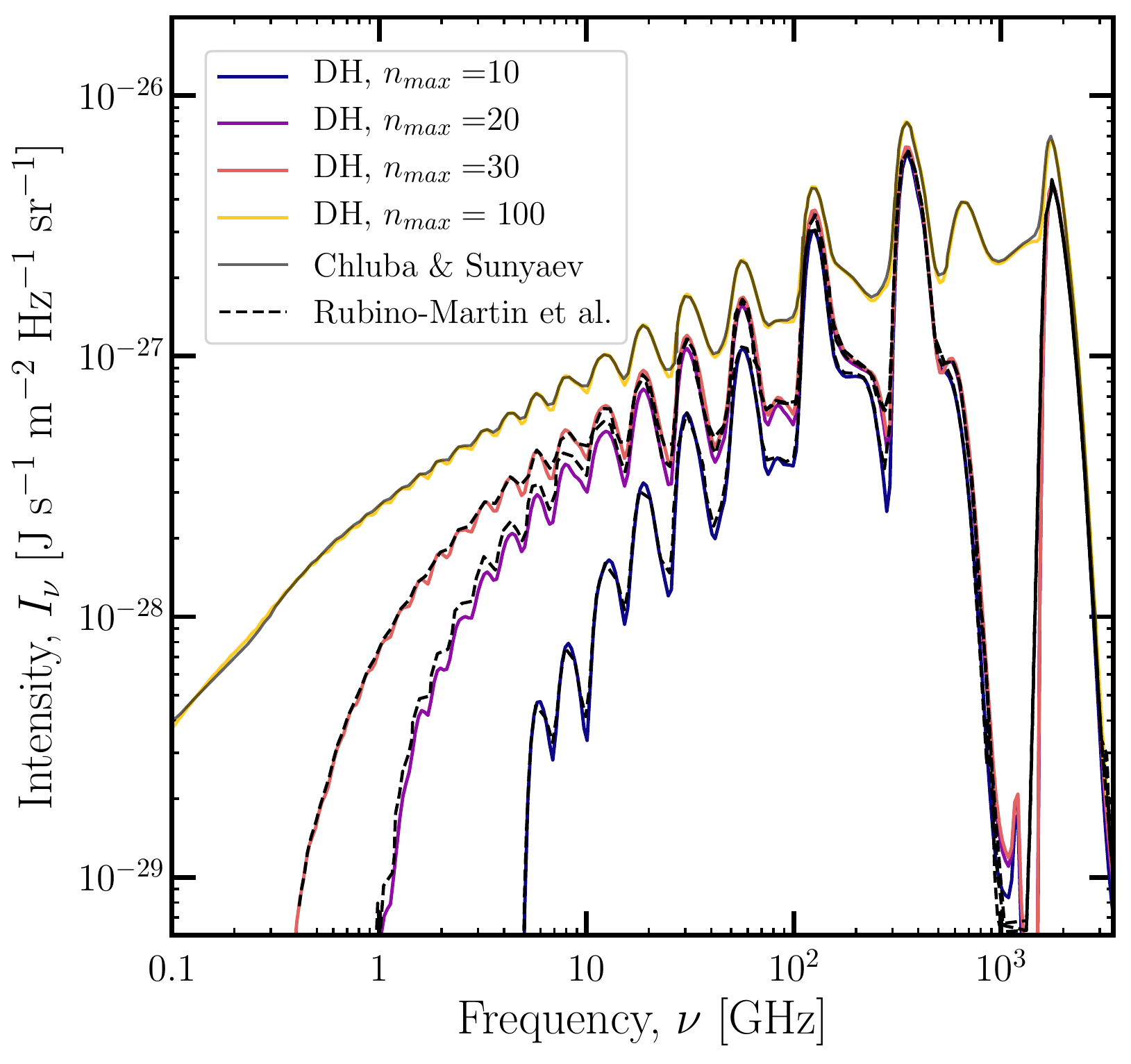}
	\caption{Spectral distortion due to atomic transitions at recombination, without exotic energy injection and tracking up to various $n_\text{max}$.
	For comparison, we also show the results of Ref.~\cite{Jens2006} in black dashed lines and Ref.~\cite{Chluba:2006xa} in the solid grey using the same values for $n_\text{max}$.}
	\label{fig:distortion_xcheck}
\end{figure}
Next, we compare our calculation of the CMB distortion due to atomic transitions to the outputs of Refs.~\cite{Jens2006, Chluba:2006xa}, calculated at different values for $n_\text{max}$.
In Fig.~\ref{fig:distortion_xcheck}, we show the hydrogen recombination spectrum calculated using our code for the same values of $n_\text{max}$ and no exotic energy injection.
For $n_\text{max} = 10$, 20, and 30, we are able to reproduce the recombination spectra from Ref.~\cite{Jens2006}, and for $n_\text{max} = 100$, we match the results from Ref.~\cite{Chluba:2006xa}.
In each case, we find only percent-level deviations compared to the results in the literature.

Lastly, the $R_{i \to j}$ transition rates are closely related to the probabilities for a hydrogen atom in the excited state to decay to the ground state with the emission of a Ly-$\alpha$ photon (e.g. by cascading through the $2p$ state, rather than the $2s$ state).
Table 1 of Ref.~\cite{Hirata:2005mz} lists these probabilities up to $n=30$, and we are able to reproduce the results to the last significant digit they report.

\section{Conclusion}
\label{sec:conclusion}

In addition to modifying the global temperature and ionization history, exotic energy injections can change the spectrum of the universe's background photons, resulting in CMB spectral distortions apart from those expected in $\Lambda$CDM.
In this work, we have described major upgrades to the \dhis code which are necessary for self-consistently tracking these photons.
We summarize the main points below:
\begin{enumerate}
	\item We extend our treatment of high-energy electrons to lower energies, which allows us to track the spectrum of secondary photons from ICS of CMB photons off low-energy electrons and simultaneously provides a more precise treatment of heating and ionization from low-energy electrons; the latter is relevant to constraints on light DM.
	
	\item We calculate the $y$-type spectral distortion caused by gas heating.
	
	\item Instead of treating hydrogen as a TLA, we now track an arbitrary number of energy levels and the subsequent line emissions from transitions between these levels. 
	
	\item We can also account for the back-reaction of the altered background spectrum of photons on these transitions.
\end{enumerate}
At each step of the upgrade, we have cross-checked our results against various other codes.
This version of \dhis is publicly available on GitHub \githubmaster.

Throughout this work, we treat energy deposition as homogeneous.
This has also been a common assumption in previous works; however, structure formation in the late universe leads to inhomogeneous energy injection and deposition~\cite{Schon:2014xoa,Schon:2017bvu}.
We leave a detailed study of these inhomogeneities to future work.

Being able to track the late time spectrum of photons paves the way for future studies of observables from exotic energy injection.
In \citetalias{paperII}, we will demonstrate various directions that can be taken with this new technology, including examining the possibility of observing spectral distortions from yet unconstrained dark matter models using future CMB spectral distortion experiments, extending existing CMB anisotropy contraints, and setting limits on couplings between ALPs and photons.
Other possible follow-ups include studying various ways in which the modified photon background could affect signals from 21-cm cosmology and the extragalactic background light.
We leave this to future work.

\section*{Acknowledgements}

We thank Sandeep Acharya and Rishi Khatri for the use of their Green's functions data, as well as Trey W. Jensen and Yacine Ali-Ha\"{i}moud for useful discussions.
TRS was supported by the Simons Foundation (Grant Number 929255, T.R.S) and by the National Science Foundation under Cooperative Agreement PHY-2019786 (The NSF AI Institute for Artificial Intelligence and Fundamental Interactions, \url{http://iaifi.org/}).
W.Q. and G.W.R. were supported by the National Science Foundation Graduate Research Fellowship under Grant No. 1745302.
W.Q. was also supported by National Science Foundation Graduate Research Fellowship under Grant No. 2141064 and G.W.R. by the U.S. Department of Energy, Office of Science, Office of Nuclear Physics under grant Contract Number DE-SC0011090.
W.Q., G.W.R., and T.R.S. were supported by the U.S. Department of Energy, Office of Science, Office of High Energy Physics of U.S. Department of Energy under grant Contract Number DE-SC0012567. 
HL is supported by the DOE under Award Number DE-SC0007968, NSF grant PHY2210498, and the Simons Foundation.

This work made use of 
\texttt{Jupyter}~\cite{Kluyver2016JupyterN}, 
\texttt{matplotlib}~\cite{Hunter:2007ouj}, 
\texttt{NumPy}~\cite{Harris:2020xlr}, 
\texttt{SciPy}~\cite{2020NatMe..17..261V}, and 
\texttt{tqdm}~\cite{daCosta-Luis2019},
as well as Webplotdigitizer~\cite{Rohatgi2022}.

\clearpage
\onecolumngrid
\appendix 

\section{Collisional Ionization and Excitation Rates}
\label{app:collisional_rates}

For collisional ionization, we adopt the results of Ref.~\cite{Kim_Rudd}, using the binary-encounter-Bethe model for HI shown in Eq.~(57)---which shows excellent agreement with experimental results between \SI{13.6}{\eV} and \SI{3}{\kilo\eV}---and the binary-encounter-dipole model for HeI and HeII, shown in Eq.~(55). All quantities required for the cross section are tabulated in Table~I of the same paper. Note that these cross-section fits are not expected to hold when the incoming electron is relativistic; however, this does not affect our results significantly, since relativistic electrons in the early universe lose their energy predominantly through ICS, with ionization being a small contribution to the total energy loss. 

For collisional excitation rates, we rely on the tabulated cross sections in Ref.~\cite{Stone_Kim_Desclaux} for hydrogen $np$ states and HeI excitation (we only track excitation up to the $2p$ state) in the energy range of \SI{10}{\eV}--\SI{3}{\kilo\eV}. For all other hydrogen states, we use the data provided by the CCC database~\cite{CCC}, which gives the cross sections of excitations from the ground state of hydrogen up to and including the $4f$ state, between \SI{14}{\eV} and \SI{999}{\eV}. 

For energies higher than those that are tabulated above, we use the Bethe approximation~\cite{RevModPhys.43.297}, which expands the excitation cross section as a function of $\mathcal{R}/E' \ll 1$ for incoming electron energies above \SI{1}{\kilo\eV}. Ref.~\cite{Stone_Kim_Desclaux} provides a nonrelativistic Bethe approximations---suitable for $E' < \SI{10}{\kilo\eV}$---for the excitation cross sections of hydrogen $np$ states and HeI excitation of the form 
\begin{alignat}{1}
    \sigma_{np} (E') = \frac{4 \pi a_0^2 \mathcal{R}}{T + B + E_{1s \to np}} \frac{f_\text{accu}}{f_\text{sc}} \left[a_{np} \log \left(\frac{E'}{\mathcal{R}}\right) + b_{np} + c_{np} \frac{\mathcal{R}}{E'}\right] \,,
\end{alignat}
where $a_0$ is the Bohr radius, $B$ is the binding energy of the ground state electron, $E_{1s \to np}$ is the excitation energy of the $np$ state, and $a_{np}$, $b_{np}$ and $c_{np}$ are fit coefficients. $f_\text{accu} / f_\text{sc}$ is 1 for hydrogen, and is a correction factor applied to atoms with more than one electron in the ground state. All unknown values in this expression are tabulated in Ref.~\cite{Stone_Kim_Desclaux}. For all other hydrogen states, we perform a fit to the three data points with the highest energies in the CCC database with the following functional form, which is appropriate for transitions which are optically forbidden~\cite{RevModPhys.43.297}: 
\begin{alignat}{1}
    \sigma_{nl}(E') = \frac{4 \pi a_0^2}{E' / \mathcal{R}} \left(\beta_{nl} + \frac{\gamma_{nl}}{E' / \mathcal{R}}\right) \,,
\end{alignat}
where $\beta_{nl}$ and $\gamma_{nl}$ are fit coefficients for each $nl$ state, and use this asymptotic form between \SI{1}{\kilo\eV} and \SI{10}{\kilo\eV}. 

Above \SI{10}{\kilo\eV}, relativistic corrections start to become important. In this regime, we switch to the relativistic version of the Bethe approximation. For optically allowed transitions, this is of the form~\cite{RevModPhys.43.297}
\begin{alignat}{1}
    \sigma_{np} = \frac{8 \pi a_0^2}{m_e \beta^2 / \mathcal{R}} \left( M_{np}^2 \left[\log \left(\frac{\beta^2}{1 - \beta^2}\right) - \beta^2\right] + C_{np} \right) \,,
\end{alignat}
where $\beta$ is the electron velocity, $M_{np}^2 = a_{np}$, $C_{np} = a_{np} \log (2 m_e \zeta_{np} / \mathcal{R})$, and $\zeta_{np} = (\mathcal{R} / 4) \exp (b_{np} / a_{np})$; $a_{np}$ and $b_{np}$ are the same coefficients used in the nonrelativistic Bethe approximation. Similarly, for optically forbidden transitions, we have
\begin{alignat}{1}
    \sigma_{nl}(E') = \frac{8 \pi a_0^2}{\mathcal{R} / (m_e \beta^2)} \beta_{nl} \,,
\end{alignat}
where once again $\beta_{nl}$ is the same coefficient as above. 

For HeII excitation, we continue using the cross section provided in Ref.~\cite{PhysRevA.55.329}.

\section{Derivation of $y$-parameter from heating}
\label{app:y_deriv}

As mentioned in Section~\ref{sec:y-type}, at early enough times, one can write the $y$-parameter as an integral over the heating rate.
Here we present the full details of the derivation.

First, one can rewrite Eq.~\eqref{eq:general_Tm_evolution} in terms of $T_m / T_\text{CMB}$ as
\begin{equation}
	\frac{d}{dt} \left( \frac{T_m}{T_\text{CMB}} \right) + H (1+J) \frac{T_m}{T_\text{CMB}} = JH + \frac{\dot{T}_m^\text{inj}}{T_\text{CMB}} .
\end{equation}
We can now write $T_m = T_m^{(0)} + \Delta T$, and note that $T_m^{(0)}$ solves the differential equation above without $\dot{T}_m^\text{inj}$, i.e. 
\begin{alignat}{1}
	\frac{d}{dt} \left(\frac{T_m^{(0)}}{T_\text{CMB}}\right) = JH \left(1 - \frac{T_m^{(0)}}{T_\text{CMB}}\right) - H \frac{T_m^{(0)}}{T_\text{CMB}} \,.
\end{alignat}
Since $J \gg 1$ prior to $1+z \approx 500$, the first term on the right-hand side drives $T_m^{(0)} \to T_\text{CMB}$ until the two terms are roughly equal; in other words, any large deviations of $T_m^{(0)}$ from $T_\text{CMB}$ are erased on a timescale much faster than the Hubble timescale, leaving the right-hand side at a value close to zero. This means that for $J \gg 1$, 
\begin{alignat}{1}
	T_m^{(0)} \approx \left(1 - \frac{1 }{J}\right) T_\text{CMB} \,.
	\label{eqn:analytic_Tm}
\end{alignat}
The temperature evolution equation also gives
\begin{alignat}{1}
	\frac{d}{dt} \left(\frac{\Delta T}{T_\text{CMB}}\right) = - H (1 + J) \left(\frac{\Delta T}{T_\text{CMB}} \right) + \frac{\dot{T}_m^\text{inj}}{T_\text{CMB}} \,.
\end{alignat}
Applying the same argument as before, we can see that $\Delta T \to 0$ until the two terms on the right-hand side are roughly equal, with any large deviations in $\Delta T$ from zero being erased well before a Hubble time. This then gives 
\begin{alignat}{1}
	\Delta T \approx \frac{\dot{T}_m^\text{inj}}{H J} = \frac{3 (1 + \chi + x_e) m_e \dot{T}_m^\text{inj}}{8 \sigma_T u_\text{CMB} x_e} \,.
\end{alignat}
for $J \gg 1$. 

With this expression, it is easy to see from Eq.~\eqref{eqn:y_DM} that the $y$-parameter due to energy injection is
\begin{align}
	y_\text{inj} &\approx \int_0^t dt\, \frac{3 n_\text{H} (1 + \chi + x_e) \dot{T}_m^\text{inj}}{8 u_\text{CMB}} = \frac{1}{4} \int_0^t dt\, \frac{\dot{Q}}{\rho_\text{CMB}} \,.
\end{align}
%

\section{Atomic Transition Rates}
\label{app:atomic_rates}

To calculate the bound-free and bound-bound transition rates, we follow the method outlined in Ref.~\cite{preHyrec}.
Starting with the bound-free rates, we calculate the recombination rate to state $nl$ and the photoionization rate from state $nl$ using~\cite{Burgess1965}
\begin{alignat}{1}
    \alpha_{nl} & = \left( \frac{2\pi}{\mu_e T_m} \right)^{3/2}
    \int_0^{\infty} e^{-\mathcal{R} \kappa^2/T_m} \gamma_{nl} [1+f^\text{CMB}+\Delta f] d(\kappa^2) , \nonumber \\
    \beta_{nl} &= \int_{\omega_{nl}}^{\mathcal{R}} d\omega [f^\text{CMB}+\Delta f] a_{nl}(k^2) \, .
    \label{eqn:bound_free_rates}
\end{alignat}
Above, $\mu_e$ is the reduced mass of the electron and proton, $\kappa = p_e a_0$ is the momentum of the unbound electron in units of the Bohr radius, $a_0$, and $\omega_{nl}$ is the energy required to photoionize a hydrogen atom in the $nl$ state.
In the expression for $\beta_{nl}$, the integral is cut off at $\mathcal{R}$ because photons with energy above this are assumed to ionize the 1s state. 
$\gamma_{nl}$ and $a_{nl}(k^2)$ are defined by
\begin{alignat}{1}
    \gamma_{nl} &= \frac{2}{3n^2} \frac{\mathcal{R}}{2\pi} (1+n^2 \kappa^2)^3 \nonumber \\
    & \;\;\;\;\; \times \sum_{l' = l\pm1}\text{max}(l,l') g(n,l,\kappa,l')^2 \nonumber \\
    a_{nl}(k^2) &= \left( \frac{4\pi \alpha a_0^2}{3}\right) n^2 (1+n^2 \kappa^2) \nonumber \\
    & \;\;\;\;\; \times \sum_{l' = l\pm1}\frac{\text{max}(l,l')}{2l+1} g(n,l,\kappa,l')^2
    \label{eqn:gamma_and_a}
\end{alignat}
$g(n,l,\kappa,l')$ is proportional to a matrix element of the dipole transition operator, 
and we calculate it using an iterative procedure described in Ref.~\cite{Burgess1965}.

Turning to the bound-bound rates, we calculate dipole up-transitions and down-transitions using
\begin{alignat}{1}
    R_{nl\to n'l'} &= A_{nl\to n'l'} [1+f^\text{CMB}+\Delta f] \;\;\;\; E_n > E_{n'} , \\
    R_{nl\to n'l'} &= \frac{g_{l'}}{g_l} A_{n'l'\to nl} [f^\text{CMB}+\Delta f] \;\;\;\; E_n < E_{n'}  \, .
    \label{eqn:bound_bound_rates}
\end{alignat}
Here, $A_{nl\to n'l'}$ is the Einstein A-coefficient, which we calculate using an iterative procedure described in Ref.~\cite{Hey_2006} (See their Eqs.~(52)-(53)), and $g_l$ or $g_i$ is the degeneracy of the corresponding energy level. 

When added to the MLA equations, Eq.~\eqref{eq:general_xi_evolution}, the rates $R_{1s\to np}$ and $R_{np\to 1s}$ need to be treated with more care. 
While it is technically true that one could solve the MLA using the $R_{1s\to np}$ rates as defined above, 
one would have to use a step size smaller than the fastest timescale, $R_{1s\to 2p}^{-1}$, 
to be able to resolve the frequent emission and absorption of Lyman-series photons. 
Instead, the standard method is to use a much larger stepsize and replace $R_{1s\to np}$ and $R_{np\to 1s}$ by smaller effective rates. 
These effective rates only keep track of the transitions that produce or absorb a Lyman-series photon that is not instantly absorbed, 
and is able to redshift out of the resonant energy line~\cite{Seager:1999km}.
To calculate these modified rates we first calculate the Sobolev optical depth and then the probability that a photon will redshift out of the resonant energy line,
\begin{alignat}{1}
    \tau_{ij} &= \frac{A_{ji} \lambda_{ij}^3 [n_i (g_j/g_i) - n_j]}{8 \pi H(z)} \\
    p_{ij} &= \frac{1-\exp (-\tau_{ij})}{\tau_{ij}}  \, .
    \label{eqn:Sobolev}
\end{alignat}
where $\lambda_{ij}$ is the line photon's wavelength.

In addition to the dipole transition rates, we include the most important quadrupole transition, $1s \leftrightarrow 2s$.
The transition rate from the $2s$ to $1s$ state in the presence of a background radiation field with occupation number $f(\omega)$ is given by~\cite{Chluba:2005uz}
\begin{equation}
	A_{2s1s} = \frac{A_0}{2} \int_0^1 \phi(y) \left[ 1 + f^\gamma (\omega) \right] \left[ 1 + f^\gamma (E_\alpha - \omega) \right] \,dy ,
	\label{eqn:A2s1s}
\end{equation}
where $A_0 =  \SI{4.3663}{\s^{-1}}$, $y = \omega / E_\alpha$, 
and $\phi (y)$ is proportional to the probability density for emitting two photons at frequencies $\omega$ and $E_\alpha - \omega$.
The factor of $1/2$ is required since by integrating $y$ from 0 to 1, we count each photon twice.
We use the analytic fit for $\phi(y)$ given in Refs.~\cite{1984A&A...138..495N,Chluba:2005uz}.
\begin{equation}
	\phi(y) = C [w(1-4^{c_3} w^{c_3}) + c_1 w^{{c_2} + {c_3}} 4^{c_3}] ,
\end{equation}
where $w = y(1-y)$, $C = 46.26$, $c_1 = 0.88$, ${c_2} = 1.53$, and ${c_3} = 0.8$.
One can check that in the absence of a radiation field, i.e. $f^\gamma (\omega) = 0$, Eq.~\eqref{eqn:A2s1s} yields the transition rate in vacuum, $A_{2s1s} = \SI{8.22}{\s^{-1}}$.
The reverse rate is
\begin{equation}
	A_{1s2s} = \frac{A_0}{2} \int_0^1 \phi(y) f^\gamma (\omega) f^\gamma (E_\alpha - \omega) \,dy .
\end{equation}

We would like to obtain the spectrum of photons resulting from this transition.
If $n_{2s}$ and $n_{1s}$ are the number of atoms in the $2s$ and $1s$ states, respectively, 
then the net change in number of photons per unit time and volume in the energy bin containing $y_i$ with width $dy_i$ is given by
\begin{align}
	n_B \frac{d N_\omega}{dt} &= 2 \,dy_i \left( n_{2s} \frac{d A_{2s1s}}{dy} - n_{1s} \frac{d A_{1s2s}}{dy} \right) \n
	&= A_0 \phi(y_i) \,dy_i \times \left\{ n_{2s} \left[ 1 + f^\gamma (\omega_i) \right] \left[ 1 + f^\gamma (E_\alpha - \omega_i) \right] - n_{1s} f^\gamma (\omega_i) f^\gamma (E_\alpha - \omega_i) \right\} .
\end{align}
The factor of 2 in the first line accounts for the fact that there will be a contribution from transitions corresponding to photons with energy $\omega_i$, as well as $E_\alpha - \omega_i$.

\section{Solving for excited states}
\label{app:matrix_method}

In Section~\ref{sec:steady_state}, we described our simplified evolution equations under the steady state approximation and calculated the rates $\tilde{\alpha}_\text{B}$
, $\tilde{\beta}_\text{B}$, and $\dot{x}_\text{inj}$ in terms of the quantity $Q_k$.
The derivation outlined there required only one matrix inversion; however, since $Q_k$ is nearly equal to the identity, calculating $1-Q_k$ to adequate precision is very slow.
Our code is based upon the following procedure, which is numerically faster.

As noted in Section~\ref{sec:steady_state}, under the steady-state approximation, determining the hydrogen level populations amounts to solving the matrix equation
\begin{alignat}{1}
	x_k = \sum_{l > 1s} M^{-1}_{kl} (b_l^{1s} + b_l^\text{rec} + b_l^\text{inj}) .
\end{alignat}
In the code, we use a slightly different normalization for $M_{kl}$ and the $b_l^i$ terms such that
\begin{gather}
	M_{kl} = \delta_{kl} - \frac{\tilde{R}_{l \rightarrow k}}{\tilde{R}_k^\text{out}} , \\
	b_l^{1s} = x_{1s} \frac{\tilde{R}_{1s \rightarrow l}}{\tilde{R}_l^\text{out}} , \quad
	b_l^\text{rec} = x_e^2 n_\text{H} \frac{\alpha_l}{\tilde{R}_l^\text{out}}, \quad
	b_l^\text{inj} = \frac{\dot{x}^\text{exc}_{\text{inj}, l}}{\tilde{R}_l^\text{out}} .
	\label{eq:M_and_b_code}
\end{gather}

Using these expressions, we can again simplify Eq.~\eqref{eq:simplified_intermediate_xe} to obtain 
\begin{alignat}{1}
\dot{x}_e = - x_e^2 n_\text{H} \tilde{\alpha}_\text{B} + x_{1s} \tilde{\beta}_\text{B} + \dot{x}_\text{inj} \,,
\end{alignat}
where we now have
\begin{align}
\tilde{\alpha}_\text{B} &= \sum_{k > 1s} \alpha_k - \beta_k M_{kl}^{-1} b_l^\text{rec} , \\
\tilde{\beta}_\text{B} &= \beta_k M^{-1}_{kl} b^{1s}_l , \\
\dot{x}_\text{inj} &= \beta_k M^{-1}_{kl} b^\text{inj}_l + \dot{x}_\text{inj}^\text{ion} \, .
\end{align}

\section{Derivation of the Three-Level Atom Model}
\label{app:derivation_TLA}

We begin with the multi-level atom model discussed in Eq.~\eqref{eq:final_xe}, given by 
\begin{alignat}{1}
    \dot{x}_e = - x_e^2 n_\text{H} \tilde{\alpha}_\text{B} + x_{1s} \tilde{\beta}_\text{B} + \dot{x}_\text{inj} \,,
    \label{eq:xe_evolution_appendix}
\end{alignat}
where we have defined the following objects:
\begin{alignat}{1}
    \tilde{\alpha}_\text{B} &= \sum_{k > 1s}  Q_k \alpha_k \,, \nonumber \\
    \tilde{\beta}_\text{B} &= \sum_{k > 1s} (1 - Q_k) \tilde{R}_{1s \to k} \,, \nonumber \\
    \dot{x}_\text{inj} &= \sum_{k > 1s} (1 - Q_k) \dot{x}_{\text{inj}, k}^\text{exc} + \dot{x}_\text{inj}^\text{ion} \,, \nonumber \\
    Q_k &= \sum_{l > 1s} M_{lk}^{-1} \tilde{R}_{l \to 1s} \,, \nonumber \\
    M_{kl} &= \delta_{kl} \tilde{R}_k^\text{out} - \tilde{R}_{l \to k} \,. 
\end{alignat}
It is also useful to recall the relation 
\begin{alignat}{1}
    \beta_k = \sum_{l > 1s} M_{lk} - \tilde{R}_{k \to 1s}
\end{alignat}

The first assumption to derive the TLA is that all $n \geq 2$ states are in Boltzmann equilibrium with each other. Under this assumption, there are no net bound-bound transitions. Furthermore, since $g_i \exp(-\omega_i / T) \tilde{R}_{i \to j} = g_j \exp(-\omega_j / T) \tilde{R}_{j \to i}$ for any two states $i$ and $j$ by detailed balance, we note that
\begin{alignat}{1}
    g_l e^{-\omega_l / T} M_{kl} = g_k e^{- \omega_k / T} M_{lk} \,, \qquad g_l e^{- \omega_l / T} M_{kl}^{-1} = g_k e^{- \omega_k / T} M_{lk}^{-1} \,.
    \label{eq:symmetry_of_M}
\end{alignat}
We also have the following detailed balance relation between photoionization and recombination coefficients, obtained using the Saha relation:
\begin{alignat}{1}
    \alpha_k = \frac{\lambda_\text{th}^3}{2} g_k e^{- \omega_k / T} \beta_k \,,
    \label{eq:thermal_eq_alpha_beta_relation}
\end{alignat}
where $\lambda_\text{th} \equiv \sqrt{2 \pi / (m_e T_\text{CMB})}$ is the thermal de Broglie wavelength of an electron. Now, revisiting the expression for $x_k$, under the assumption of Boltzmann equilibrium for all excited states, we can write with the help of Eqs.~\eqref{eq:symmetry_of_M} and~\eqref{eq:thermal_eq_alpha_beta_relation} 
\begin{alignat}{1}
    x_k &= \sum_{l > 1s} M_{kl}^{-1} \left( x_{1s} \tilde{R}_{1s \to l} + x_e^2 n_\text{H} \alpha_l + \dot{x}_{\text{inj}, l}^\text{exc} \right) \nonumber \\
    &= \sum_{l > 1s} M_{kl}^{-1} \left( x_{1s} \tilde{R}_{1s \to l} + x_e^2 n_\text{H} \frac{\lambda_\text{th}^3}{2} g_l e^{-\omega_l / T} \beta_l + \dot{x}_{\text{inj},l}^\text{exc} \right) \nonumber \\
    &= \sum_{l > 1s} M_{kl}^{-1} \left[ \frac{x_{1s}}{2} e^{\omega_1/T} g_l e^{-\omega_l / T} \tilde{R}_{l \to 1s} + x_e^2 n_\text{H} \frac{\lambda_\text{th}^3}{2} g_l e^{- \omega_l / T} \left( \sum_{p > 1s} M_{pl} - \tilde{R}_{l \to 1s} \right) + \dot{x}_{\text{inj},i}^\text{exc} \right] \nonumber \\
    &= \frac{1}{2} \left(x_{1s} e^{\omega_1 / T} - x_e^2 n_\text{H} \lambda_\text{th}^3 \right) \sum_{l > 1s}  g_l e^{- \omega_l / T} M_{kl}^{-1} \tilde{R}_{l \to 1s} + \frac{1}{2} x_e^2 n_\text{H} \lambda_\text{th}^3 \sum_{l > 1s} \sum_{q > 1s} g_l e^{- \omega_l / T} M_{kl}^{-1} M_{ql} + \sum_{l > 1s} M_{kl}^{-1} \dot{x}_{\text{inj}, l}^\text{exc} \nonumber \\
    &\approx \frac{1}{2} \left(x_{1s} e^{\omega_1 / T} - x_e^2 n_\text{H} \lambda_\text{th}^3 \right) g_k e^{- \omega_k / T} \sum_{l > 1s}   M_{lk}^{-1} \tilde{R}_{l \to 1s} + \frac{1}{2} x_e^2 n_\text{H} \lambda_\text{th}^3 g_k e^{- \omega_k / T} \sum_{l > 1s} \sum_{q > 1s} M_{lk}^{-1} M_{ql} + \sum_{l > 1s} M_{kl}^{-1} \dot{x}_{\text{inj}, l}^\text{exc} \nonumber \\
    &= \frac{1}{2} g_k e^{- \omega_k / T} \left[ \left(x_{1s} e^{\omega_1/T} - x_e^2 n_\text{H} \lambda_\text{th}^3 \right) Q_k + x_e^2 n_\text{H} \lambda_\text{th}^3 \right] + \sum_{l > 1s} M_{kl}^{-1} \dot{x}_{\text{inj},l}^\text{exc} \,. 
\end{alignat}
At this point, in order to derive expressions previously used in the literature, we must make the further approximation that $\dot{x}_{\text{inj},l}^\text{exc}$ is small, and can be dropped; we will return to this point later in the section. Doing so, and applying the assumption about a Boltzmann distribution of states, we find
\begin{alignat}{1}
    \frac{g_k e^{-\omega_k / T}}{2 e^{- \omega_2 / T}} x_{2s} \approx \frac{1}{2} g_k e^{- \omega_k / T} \left[ \left(x_{1s} e^{\omega_1/T} - x_e^2 n_\text{H} \lambda_\text{th}^3 \right) Q_k + x_e^2 n_\text{H} \lambda_\text{th}^3 \right] \,,
\end{alignat}
or
\begin{alignat}{1}
    Q_k \approx \frac{x_{1s} e^{\omega_1/T} - x_e^2 n_\text{H} \lambda_\text{th}^3}{x_{2s} e^{\omega_2 / T} - x_e^2 n_\text{H} \lambda_\text{th}^3} \equiv Q \,.
\end{alignat}
Note that the right-hand side of this expression is independent of $k$, i.e.\ $Q_k$ takes the same approximate value $Q$ for all states $k$. 

In fact, $Q$ can be written purely in terms of atomic transition rates, independent of the population of the various hydrogen states and the free electron fraction. Beginning from the definition of $Q_k$, we can write the sum
\begin{alignat}{1}
    \sum_{l > 1s} \sum_{k > 1s}  g_k e^{-\omega_k / T}  M_{lk} Q_k &= \sum_{l > 1s} \sum_{k > 1s}  g_k e^{-\omega_k / T} M_{lk} \sum_{p > 1s} M_{pk}^{-1} \tilde{R}_{p \to 1s} \nonumber \\
    &\approx \sum_{l > 1s} \sum_{k > 1s} g_l e^{- \omega_l / T} M_{kl} \sum_{p > 1s} M_{pk}^{-1} \tilde{R}_{p \to 1s} \nonumber \\
    &= \sum_{l > 1s} g_l e^{-\omega_l / T} \tilde{R}_{l \to 1s} 
\end{alignat}
On the other hand, 
\begin{alignat}{1}
    \sum_{l > 1s} \sum_{k > 1s} g_k e^{-\omega_k / T} M_{lk} Q_k &= \sum_{k > 1s} g_k e^{-\omega_k / T} Q_k \sum_{l > 1s} M_{lk} \nonumber \\
    &= \sum_{k > 1s} g_k e^{-\omega_k / T} Q_k (\beta_k + \tilde{R}_{k \to 1s})
\end{alignat}
Given the fact that $Q_k$ is approximately constant, we can put these two expression together to find
\begin{alignat}{1}
    Q \approx \frac{\sum_{k > 1s} g_l e^{-\omega_k / T} \tilde{R}_{k \to 1s}}{\sum_{k > 1s} g_k e^{-\omega_k / T} \beta_k + \sum_{k > 1s} g_k e^{-\omega_k / T} \tilde{R}_{k \to 1s}} \,.
\end{alignat}
The final approximation that we make is that recombination to the ground state from states $n > 2$ is negligible compared to the $n = 2$ states, which is an excellent approximation assuming the excited states are Boltzmann distributed~\cite{preHyrec}. Defining the case-B photoionization rate $\beta_\text{B} = (1/8)\exp(\omega_2 / T) \sum_{k > 1s} g_k \exp(-\omega_k / T) \beta_k$, this gives us our final expression for $Q$, 
\begin{alignat}{1}
    Q &\approx \frac{2 e^{-\omega_2 / T} (\tilde{R}_{2s \to 1s} + 3 \tilde{R}_{2p \to 1s}) }{8 e^{-\omega_2 / T} \beta_\text{B} +  2 e^{-\omega_2 / T} (\tilde{R}_{2s \to 1s} + 3 \tilde{R}_{2p \to 1s})} = \frac{\tilde{R}_{2s \to 1s} / 4 + 3 \tilde{R}_{2p \to 1s} / 4}{\beta_\text{B} + \tilde{R}_{2s \to 1s} / 4 + 3 \tilde{R}_{2p \to 1s} / 4} = \mathcal{C} \,,
\end{alignat}
where in the last step we note that the expression is exactly to the Peebles-$\mathcal{C}$ factor. With this approximation, we find
\begin{alignat}{1}
    \tilde{\alpha}_\text{B} &\approx Q \sum_{k > 1s} \alpha_k = \mathcal{C} \alpha_\text{B} \,, \nonumber \\
    \tilde{\beta}_\text{B} &\approx (1 - \mathcal{C}) \sum_{k > 1s} \frac{g_k e^{-\omega_k / T}}{2 e^{-\omega_1 / T}} \tilde{R}_{k \to 1s} \approx (1 - \mathcal{C}) e^{-(\omega_2 - \omega_1)/T} (\tilde{R}_{2s \to 1s} + 3 \tilde{R}_{2p \to 1s}) = 4 \mathcal{C} e^{-(\omega_2 - \omega_1) / T} \beta_\text{B} \,, \nonumber \\
    \dot{x}_\text{inj} &\approx (1 - \mathcal{C}) \sum_{k > 1s}\dot{x}_{\text{inj}, k}^\text{exc} + \dot{x}_\text{inj}^\text{ion} \,,
\end{alignat}
where $\alpha_\text{B}$ is the case-B recombination coefficient. Upon substitution into Eq.~\eqref{eq:xe_evolution_appendix}, this finally leads to
\begin{alignat}{1}
    \dot{x}_e = \mathcal{C} \left(- x_e ^2 n_\text{H} \alpha_\text{B} + 4 \beta_\text{B} e^{-(\omega_2 - \omega_1) / T} x_{1s} \right) + (1 - \mathcal{C}) \sum_{k > 1s} \dot{x}_{\text{inj}, k}^\text{exc} + \dot{x}_\text{inj}^\text{ion} \,,
\end{alignat}
the evolution equation for the three-level atom. In summary, the assumptions made are that \textit{1)} all excited states follow a Boltzmann distribution with respect to each other, \textit{2)} injected photons play a small role in setting the occupation number of the excited states, \textit{3)} recombination proceeds primarily through the $n = 2$ states, which is a consequence of the Boltzmann suppression of the population in higher level states made in the first assumption. Of these assumptions, the second assumption is exactly true for standard recombination, but may be false in the presence of exotic energy injection---in fact, injected low-energy photons that excite hydrogen atoms can easily dominate the spectrum, since the CMB blackbody distribution is exponentially suppressed for such energies. With the proper inclusion of injected photons, $Q_k$ becomes $k$-dependent, and the effective recombination rate after recombining to state $k$ is no longer independent of $k$, and cannot be simply expressed as a single factor $\mathcal{C}$; the usual way of including excitations through a term proportional to $(1 - \mathcal{C})$~\cite{Chen:2003gz,DH} is therefore not correct. Of course, the presence of nonthermal photons already breaks the first assumption to begin with, necessitating the full multi-level treatment described in Section~\ref{sec:evolution}.

\section{Additional Cross Checks}
\label{app:cross_checks}

In this appendix, we validate our improvements to the \dhis code by comparing against results in the literature.

\subsection{Comparison to \dhis \texttt{v1.0}}

%
\begin{figure*}
	\includegraphics[width=\textwidth]{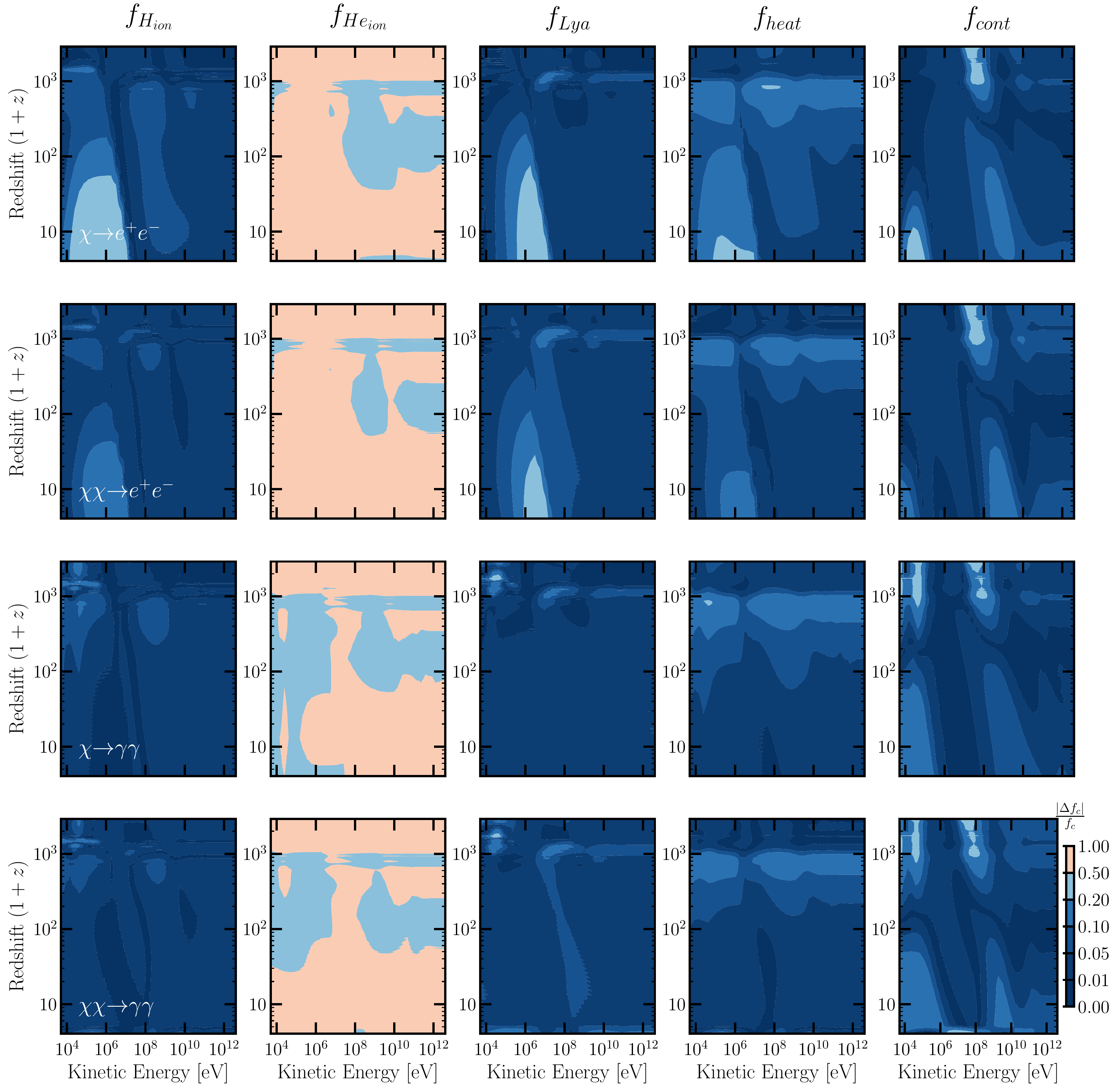}
	\caption{Comparison against the $f_c$ values calculated from \dhis \texttt{v1.0}.
		Each row shows either decay or annihilation to $e^+ e^-$ or photon pairs, and each column shows a different energy deposition channel.}
	\label{fig:f_heatmap}
\end{figure*}
In Fig.~\ref{fig:f_heatmap}, we show the difference between the $f_c$'s calculated using the updated treatment of low-energy electrons (using $n_\text{max}=10$ and only one iteration) and \dhis \texttt{v1.0}, as a function of redshift and the kinetic energy of the injected electron.
The channels are defined as follows.
\begin{itemize}
	\item H ion: energy deposited by photoionization and collisional ionization of hydrogen.
	
	\item He ion: energy deposited by photoionization and collisional ionization of helium.
	
	\item Ly-$\alpha$: energy deposited by photons and electrons into $1s \rightarrow 2p$ excitations. This was previously labeled as the `exc' channel, since this was the only excitation that we tracked.
	In the new method, since we can track an arbitrary number of excited states, then the number of Lyman-$\alpha$ photons emitted depends on the probability that an excited states cascades to the ground state by first deexciting to $2p$; as mentioned in Section~\ref{sec:atom-cross-checks}, this can be calculating using the $R_{i \to j}$ transition rates.
	
	\item heat: energy deposited by injected electrons into internal energy of the IGM.
	
	\item cont: energy deposited into photons with energy less than $E_\alpha$.
	Again, since the new method can track excited states other than $2p$, the continuum channel includes contributions from deexcitations of excited states to $2p$, as well as deexcitations to $2s$ and the two photon transition from $2s$ to $1s$.
\end{itemize}
With the exception of helium ionization and certain regions of continuum deposition, the difference in $f_c$ for all the channels is under 10\% for most redshifts and energies.
As discussed in Section~\ref{sec:lowengelec}, the helium ionization channel in MEDEA is rather noisy due to the Monte Carlo procedure they employ; this explains the large relative difference in $f_{\text{He ion}}$.

We also expect some differences in $f_\text{cont}$ since we are including a new contribution: the spectrum of photons upscattered by ICS off low-energy electrons and photoionized electrons. 
This new component is important at high redshifts, which is also where we find the largest discrepancies.
There are also a number of larger discrepancies at low redshifts in channels which primarily produce $e^+ e^-$ pairs; these are due to the fact that we are including new excitation states and are using different cross-sections from before.
We find that if we set these contributions to zero, agreement with \dhis \texttt{v1.0} is restored at the level of about 10\%.

In addition, we include an option to calculate an `effective' $f_\text{exc}$ such that if one uses the \dhis \texttt{v1.0} TLA evolution equation,
\begin{gather}
\dot{x}_e = - \mathcal{C} \left[ n_\text{H} x_e x_\text{HII} \alpha_\text{B} - 4 (1 - x_\text{HII}) \beta_\text{B} e^{-E_{21} / T_\text{CMB}} \right] + \left[ \frac{f_\text{ion}}{\mathcal{R} n_\text{H}} + \frac{f_\text{exc}}{E_\alpha n_\text{H}} \right] \left( \frac{dE}{dV \, dt} \right)^\text{inj} + \dot{x}^\text{re} ,
\label{eq:TLA_xe}
\end{gather}
together with the newly defined $f_\text{exc}$, then one obtains the same histories as that calculated using the MLA.
Note that compared to the Eq. (5) in Ref.~\cite{DH}, we have absorbed a factor of $(1-\mathcal{C})$ into $f_\text{exc}$ for numerical stability.

To summarize, with this version of \texttt{DarkHistory}, we include the option \texttt{elec\_method} in \texttt{main.evolve()}, which allows one to calculate $f_c$ by one of three methods depending on if the option is set to \texttt{`old'}, \texttt{`new'}, or \texttt{`eff'}.
\begin{itemize}
	\item \texttt{`old'}: calculate the $f_c$'s as in \dhis \texttt{v1.0}, using MEDEA for the energy deposition of electrons with energy $<3$ keV.
	
	\item \texttt{`new'}: calculate the $f_c$'s without separating low-energy electrons from high-energy electrons.
	
	\item \texttt{`eff'}: calculate the $f_c$'s without separating low-energy electrons from high-energy electrons, and also output an effective $f_\text{exc}$ that can be plugged into the TLA equations.
\end{itemize}

\subsection{Spectral distortions from ICS and heating}

As mentioned in Section~\ref{sec:y-validation}, a key difference between our work and Ref~\cite{Acharya:2018iwh} is that they treat the IGM as completely ionized.
At the early redshifts they are interested in, this is a good assumption.
However, between the redshift of $1+z=3000$ (the default initial redshift for \dhis) and recombination, there is a small but non-negligible amount of neutral hydrogen.
Secondary electrons resulting from ionization of this hydrogen can contribute to heating and therefore contribute to a $y$-type distortion.
\begin{figure*}
	\includegraphics[width=\textwidth]{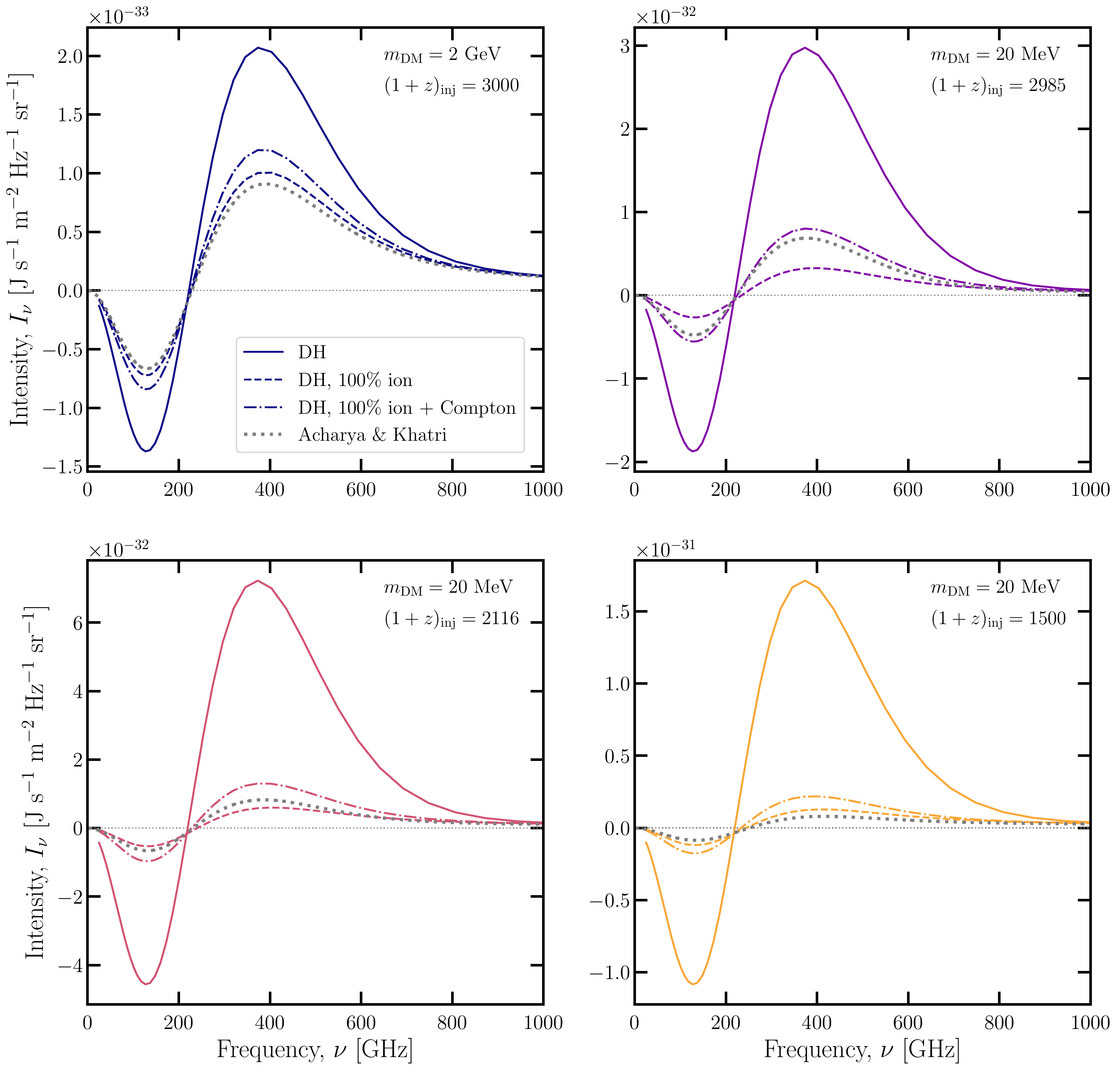}
	\caption{Spectral distortions from dark matter decaying to $e^+ e^-$ pairs, at various dark matter masses and injection redshifts. The solid lines shows the distortions contributed by low-energy photons in \texttt{DarkHistory}; the dashed lines neglect energy deposition by secondary low-energy electrons, which may result from photoionizations; the dot-dashed curves make the same assumptions as the dashed, but furthermore include the effects of heating from Compton scattering (see text for details). For comparison, the Green's functions from Ref.~\cite{Acharya:2018iwh} are shown in the gray dashed line.
	The dashed and dot-dashed curves bracket, or nearly bracket the Green's functions; this is as expected, since Ref.~\cite{Acharya:2018iwh} assumes 100\% ionization and accounts for the heating component.}
	\label{fig:greens_fncs}
\end{figure*}

Fig.~\ref{fig:greens_fncs} shows the Green's functions from Ref~\cite{Acharya:2018iwh} for dark matter of different masses decaying to $e^+ e^-$ pairs at redshift $1+z_\text{inj}$, compared to the same distortions generated by \dhis using a few different methods.
The solid line shows the component from only summing over the low-energy photon spectra; we see that as a consequence of tracking the ionization level consistently in \texttt{DarkHistory}, the solid line is always larger in amplitude than the Green's functions of Ref.~\cite{Acharya:2018iwh}-- that is to say, assuming full ionization may underestimate spectral distortions from the epoch prior to recombination.

To check consistency with Ref.~\cite{Acharya:2018iwh} under matched assumptions, we can try to turn off the contribution to heating from secondary electrons produced by ionization.
The dashed line shows the predictions of \dhis without including the heating from ionized secondary electrons in the module for low energy deposition.
While modifying the low energy deposition accounts for most of the energy going into heat, this does not fully bracket the contribution from photoionization.
Our high energy deposition transfer functions do not extend to a value of $x_e = 1$, since at the highest redshifts we consider, there is still a small but non-negligible amount of neutral hydrogen.
Hence, for this cross-check, we cannot truly set $x_e = 1$, so some ionizations are necessarily included and the secondary electrons from this channel can propagate into the low-energy electrons and contribute to heating.
Hence, while the dashed curves are lower than the Green's function from Ref~\cite{Acharya:2018iwh} for $m_\text{DM} = 20$ MeV and injection redshifts 2985 and 2116, they are slightly above for the other panels.

Finally, this comparison also omits a contribution that {\it is} included in Ref.~\cite{Acharya:2018iwh}, where photons heat the free electrons through Compton scattering (since they are not absorbed via photoionization).
Thus, the dot-dashed curves include this heating from Compton scattering, calculated using Equation (B.2) in Ref.~\cite{Acharya:2018iwh}.
These and the dashed curves fully bracket or nearly bracket the Green's functions from Ref~\cite{Acharya:2018iwh} in Fig.~\ref{fig:greens_fncs}.

Thus, we achieve reasonable agreement with Ref.~\cite{Acharya:2018iwh} when we do not include photoionizations from low-energy photons, but do include Compton scattering from these photons that would realistically photoionize at these redshifts.

\subsection{MLA treatment validation}
\label{sec:MLA_vs_TLA}

\begin{figure}[t]
	\includegraphics[width=0.5\textwidth]{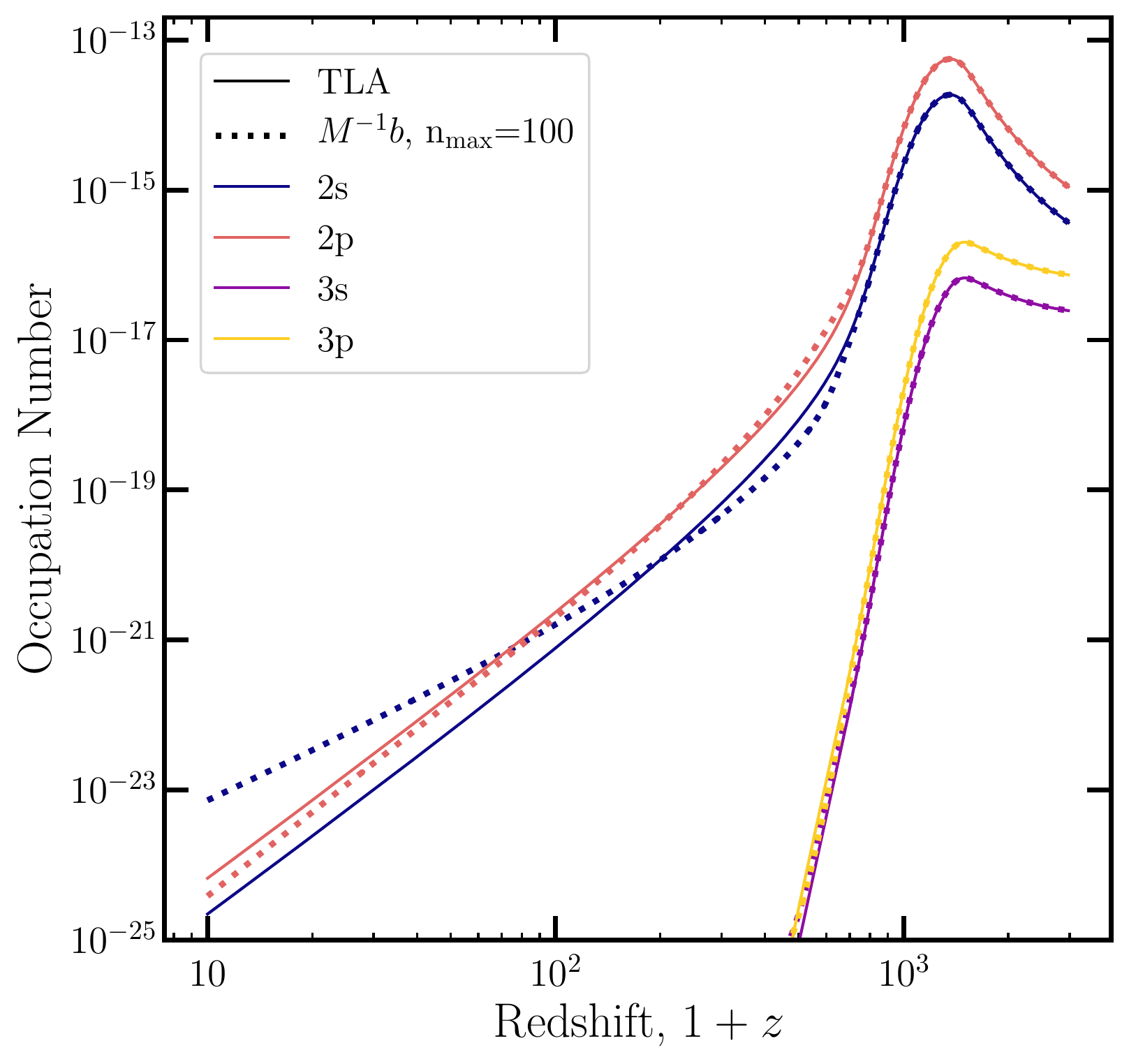}
	\caption{Occupation number of hydrogen levels under both the MLA and TLA treatments.}
	\label{fig:MLA_vs_TLA}
\end{figure}

At high enough redshifts, when the density of hydrogen is large enough that the TLA assumptions hold, the TLA and MLA treatments should yield the same results.
Fig.~\ref{fig:MLA_vs_TLA} shows the evolution of the occupation levels for the lowest hydrogen levels calculated using Eq.~\eqref{eq:matrix_MLA} and with the TLA method; both are calculated without including any sources of exotic energy injection.
The two methods agree well at redshifts above a few hundred.

\bibliography{bib}

\end{document}